\documentclass[namedreferences]{solarphysics}

\usepackage[hyperref,optionalrh]{spr-sola-addons} 
\usepackage{graphicx}        
\usepackage{color}           
\usepackage{breakurl}        

\chardef\us=`\_

\begin{document}

\begin{article}
\begin{opening}

\title{Upflows in the Upper Solar Atmosphere}

\author[corref,addressref={aff1,aff2},email={huitian@pku.edu.cn}]{\inits{H.}\fnm{Hui}\,\lnm{Tian}\orcid{0000-0002-1369-1758}}
\author[addressref={aff3,aff4}]{\inits{L.}\fnm{Louise}\,\lnm{Harra}\orcid{}}
\author[addressref=aff5]{\inits{D.}\fnm{Deborah}\,\lnm{Baker}\orcid{}}
\author[addressref=aff6]{\inits{D.}\fnm{David H.}\,\lnm{Brooks}\orcid{}}
\author[addressref=aff7]{\inits{L.}\fnm{Lidong}\,\lnm{Xia}\orcid{}}
\address[id=aff1]{School of Earth and Space Sciences, Peking University, Beijing 100871, China}
\address[id=aff2]{Key Laboratory of Solar Activity, National Astronomical Observatories, Chinese Academy of Sciences, Beijing 100012, China}
\address[id=aff3]{PMOD/WRC, Dorfstrasse 33, 7260 Davos Dorf, Switzerland}
\address[id=aff4]{ETH-Z\"urich, H\"onggerberg Campus, Z\"urich, Switzerland}
\address[id=aff5]{Mullard Space Science Laboratory, University College London, Holmbury, St. Mary, Dorking, Surrey, KT22 9XF, UK}
\address[id=aff6]{College of Science, George Mason University, 4400 University Drive, Fairfax, VA 22030, USA}
\address[id=aff7]{Shandong Provincial Key Laboratory of Optical Astronomy and Solar-Terrestrial Environment, Institute of Space Sciences, Shandong University, Weihai, 264209 Shandong, China}

\runningauthor{H. Tian et al.}
\runningtitle{Upflows}

\begin{abstract}
Spectroscopic observations at extreme- and far-ultraviolet wavelengths have revealed systematic upflows in the solar transition region and corona. These upflows are best seen in the network structures of the quiet Sun and coronal holes, boundaries of active regions, and dimming regions associated with coronal mass ejections. They have been intensively studied in the past two decades because they are likely to be closely related to the formation of the solar wind and heating of the upper solar atmosphere. We present an overview of the characteristics of these upflows, introduce their possible formation mechanisms, and discuss their potential roles in the mass and energy transport in the solar atmosphere. Although past investigations have greatly improved our understanding of these upflows, they have left us with several outstanding questions and unresolved issues that should be addressed in the future. New observations from the \textit{Solar Orbiter} mission, the \textit{Daniel K. Inouye Solar Telescope} and the \textit{Parker Solar Probe} will likely provide critical information to advance our understanding of the generation, propagation, and energization of these upflows. 
\end{abstract}
\keywords{Active Regions, Velocity Field; Coronal Holes; Coronal Mass Ejections, Low Coronal Signatures; Heating, Coronal; Spectral Line, Broadening}
\end{opening}

\section{Introduction}
     \label{S-Introduction} 

The upper solar atmosphere consists of the corona and transition region (TR), spanning a temperature range from about 3$\times$10$^{4}$ to several million Kelvin. In such a hot environment, atoms are often highly ionized and produce hundreds of strong emission lines mainly at extreme-ultraviolet (EUV) and far-ultraviolet (FUV) wavelengths ($\approx$100\,--\,1700\,{\AA}). In the past quarter century, dedicated observations with several EUV/FUV spectrographs, particularly the \textit{EUV Imaging Spectrometer} (EIS: \citealp{Culhane2007}) onboard \textit{Hinode} \citep{Kosugi2007}, the \textit{Interface Region Imaging Spectrograph} (IRIS: \citealp{DePontieu2014}), the \textit{Solar Ultraviolet Measurements of Emitted Radiation} (SUMER: \citealp{Wilhelm1995,Lemaire1997}), and the \textit{Coronal Diagnostic Spectrometer} (CDS: \citealp{Harrison1995}) onboard the \textit{Solar and Heliospheric Observatory} \citep[SOHO:][]{Domingo1995}, have greatly improved our understanding of various types of dynamic activity in the upper solar atmosphere.    

Signatures of outflows or upflows, i.e. blue shifts of spectral lines, have been frequently reported from spectroscopic observations with these facilities. In this review, we focus on the systematic and pervasive upflows observed primarily with the aforementioned spectrographs. Sporadic upflows/outflows such as coronal mass ejections and coronal jets are not discussed in this review. These systematic upflows have been one focus of investigation in the past two decades, mainly because they are likely to be related to the formation of the solar wind. In addition, these upflows are expected to contribute to the mass and energy transport in the solar atmosphere, and thus may play an important role in coronal heating. 

Here we present a review on both observational and theoretical investigations of these upflows. We first discuss upflow signatures in the quiet Sun and coronal holes in Section\,\ref{qsch}, then provide a detailed introduction to the coronal-upflow phenomenon at active-region (AR) boundaries in Section\,\ref{ar}. Section\,\ref{dimming} describes characteristics of the upflows identified from regions of coronal dimmings induced by coronal mass ejections (CMEs). Finally, we summarize the major findings about these upflows and briefly discuss future perspectives in Section\,\ref{summary}.

\section{Upflows from the Quiet Sun and Coronal Holes} 
      \label{qsch}      

\subsection{Quiet-Sun Regions}
      \label{qs} 
   
\begin{figure}    
   \centerline{\includegraphics[width=0.99\textwidth,clip=]{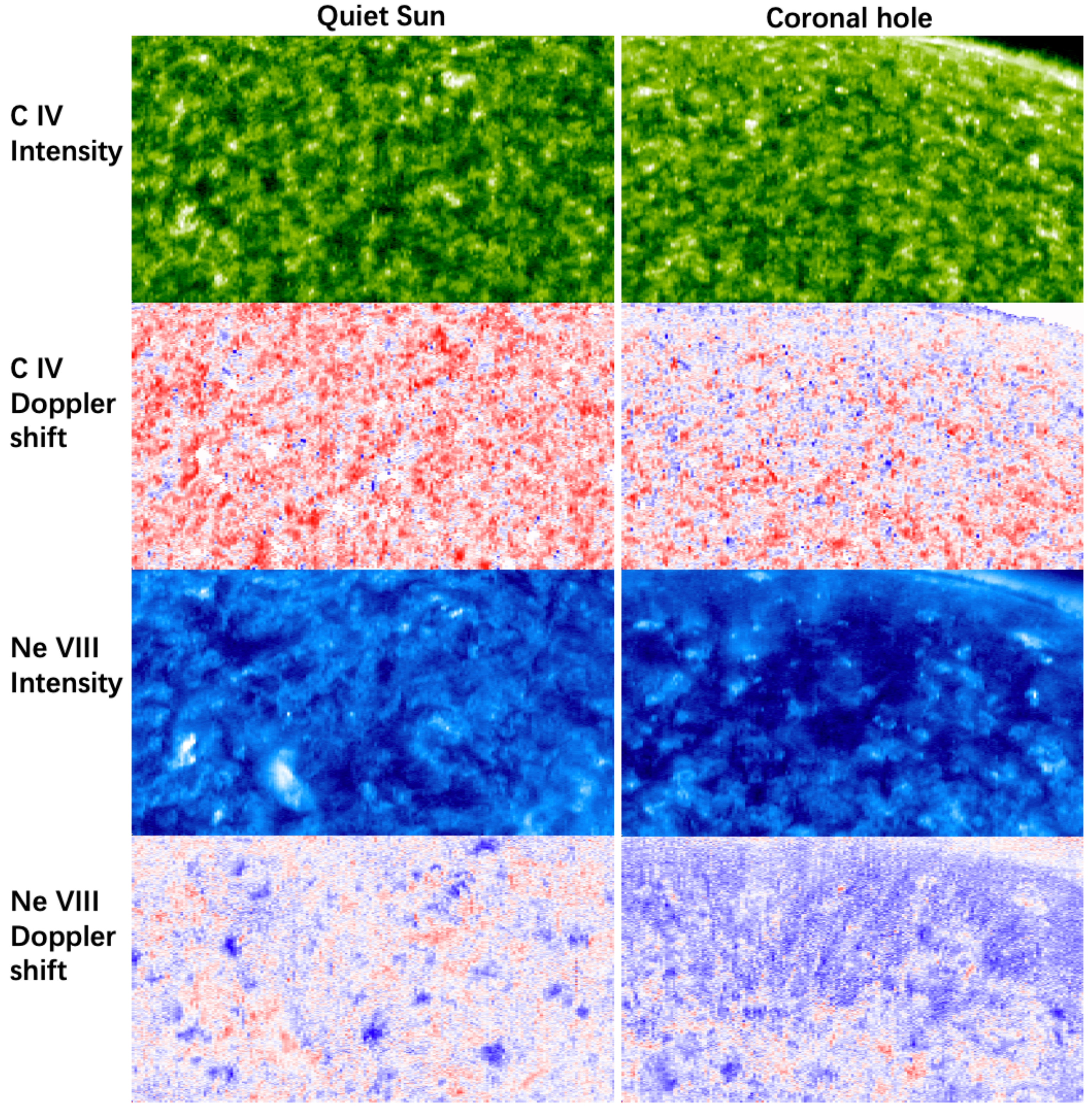}
              }
              \caption{Intensity and Doppler shift images of C\,{\sc{iv}}\,1548\,{\AA}\,and Ne\,{\sc{viii}}\,770\,{\AA}. Blue and red colors in the Dopplergrams indicate blue shifts and red shifts, respectively. These images correspond to a field of view (FOV) of 540$^{\prime\prime}$$\times$300$^{\prime\prime}$, and are taken from the SUMER Image Database: http://www2.mps.mpg.de/projects/soho/sumer/text/s029601.html. The same datasets have been analyzed by \cite{Dammasch1999} and \cite{Hassler1999}. }
   \label{fig_qsch_map}
   \end{figure}

\begin{figure}    
  \centering
\begin{minipage}[t]{0.9\textwidth}
{\includegraphics[width=\textwidth]{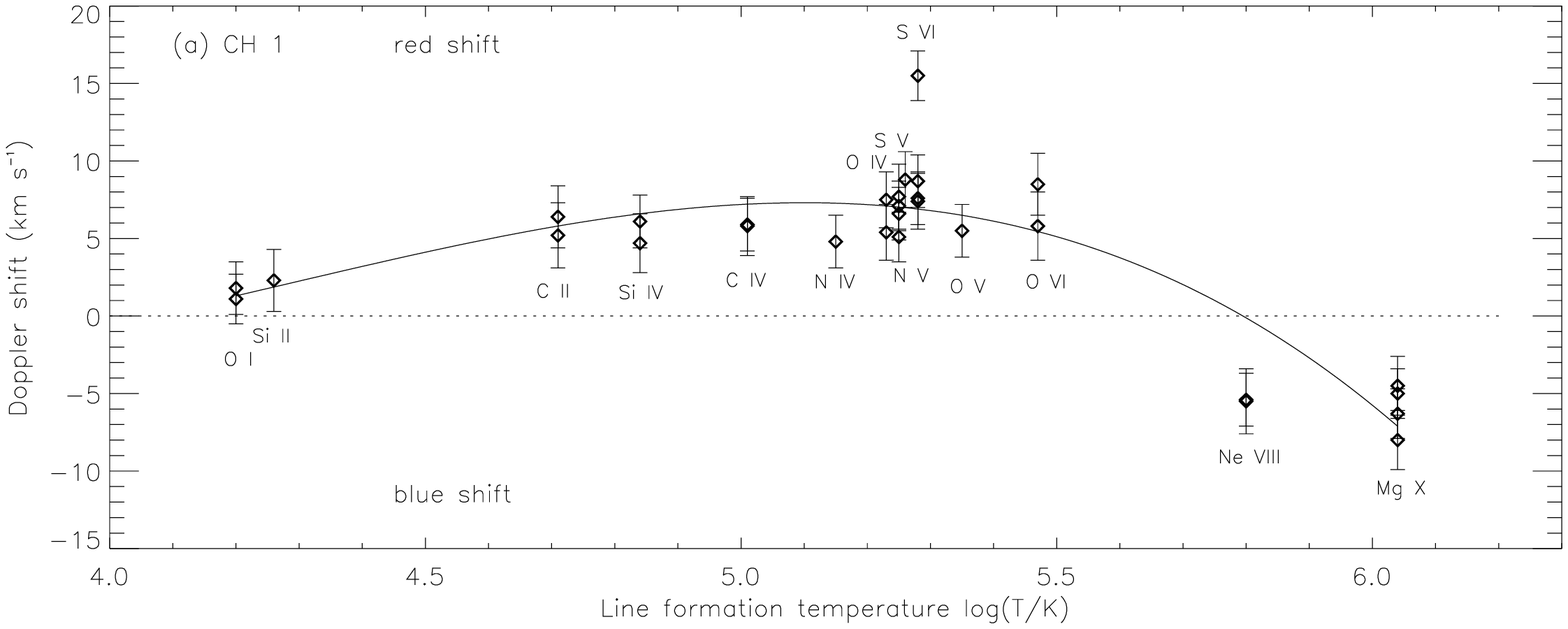}}
\end{minipage}
\begin{minipage}[t]{0.9\textwidth}
{\includegraphics[width=\textwidth]{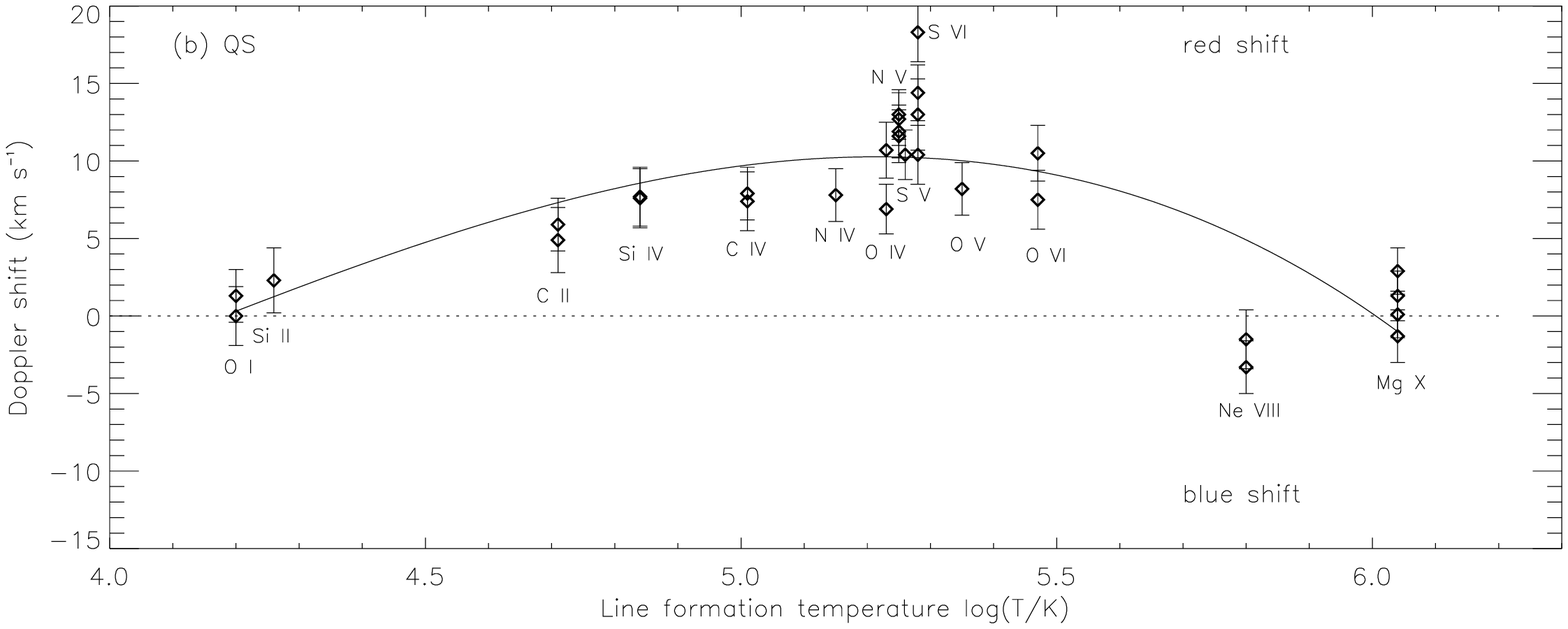}}
\end{minipage}
\begin{minipage}[t]{0.9\textwidth}
{\includegraphics[width=\textwidth]{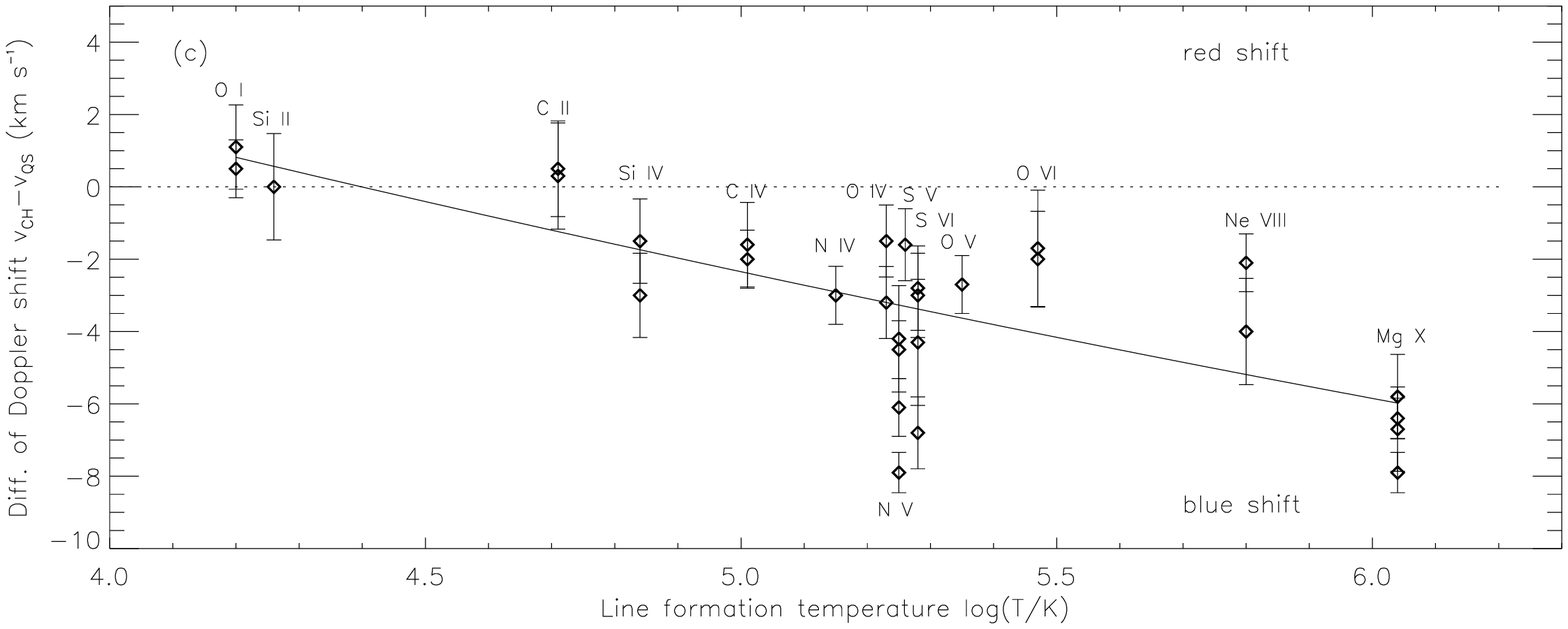}}
\end{minipage}
    \caption{Average Doppler shifts of several spectral lines with different formation temperatures \citep{Xia2004}. (a) A coronal hole region. (b) A quiet-Sun region. (c) Difference of the Doppler shifts between the coronal hole and quiet-Sun regions. Reproduced with permission from \textit{Astronomy \& Astrophysics}, $\copyright$ ESO.}
   \label{fig_qsch_curve}
   \end{figure}

It has been known since the 1970s that emission lines formed in the TR, such as C\,{\sc{iv}}\,1548\,{\AA}\,formed at a temperature of $\approx$10$^{5}$\,K under ionization equilibrium, are redshifted by a few km\,s$^{-1}$ on average in the quiet Sun \citep[e.g.\,][]{Doschek1976}. The red shift is obviously larger in network regions compared to internetwork regions (Figure\,\ref{fig_qsch_map}). With the capability of observing hundreds of strong emission lines formed in a wide range of temperatures, the SUMER spectrometer allowed detailed investigations of the Doppler shift as a function of the line-formation temperature. SUMER observations revealed a clear dependence of the Doppler shift on temperature, i.e. the average red shift increases with temperature and peaks around 2$\times$10$^{5}$\,K (or $\log\,T=5.3$) \citep[e.g.\,][]{PeterJudge1999,Stucki2000,Dadashi2011}. As the temperature continues to increase, the average red shift decreases and turns into a blue shift above $\approx$5$\times$10$^{5}$\,K. This trend can be clearly seen from Figure\,\ref{fig_qsch_curve}.
  
The change from downflow (i.e. red shifts of spectral lines) dominance to upflow dominance was unambiguously established from observations of the strong Ne\,{\sc{viii}}\,770\,{\AA}\,line, which is formed at $\approx$6.5$\times$10$^{5}$\,K in the upper TR or lower corona. Under the assumption of zero average Doppler shift above the solar limb, \cite{Dammasch1999} determined a rest wavelength of 770.428 $\pm$ 0.003\,{\AA} for this line. The same rest wavelength was also found independently by \cite{PeterJudge1999}. Using this rest wavelength, the Ne\,{\sc{viii}}\,770\,{\AA}\,line was found to be blueshifted by $\approx$2\,km\,s$^{-1}$ on average \citep[e.g.\,][]{PeterJudge1999,Teriaca1999,Xia2004}. Dopplergrams of Ne\,{\sc{viii}}\,770\,{\AA}\,obtained through raster scans revealed very prominent blue shifts at junctions of multiple adjacent network structures \citep[e.g.\,][]{Hassler1999,Dammasch1999,Aiouaz2008,Tian2008}. These localized blue shifts could reach 5\,--\,10 km\,s$^{-1}$ (Figure\,\ref{fig_qsch_map}). They generally correspond to the legs of magnetic loops reconstructed through force-free magnetic-field extrapolations, indicating mass supply to the corona along flux tubes rooted in the chromospheric network \citep{Tian2009}. Using the Doppler shift of Ne\,{\sc{viii}} as a proxy for the plasma bulk flow (i.e.\, the proton flow), these authors also found an anti-correlation between the flux tube expansion factor and mass flux.

\subsection{Coronal Holes}
      \label{ch}    
   
A similar dependence of Doppler shift on formation temperature has also been found in coronal holes. From Figure\,\ref{fig_qsch_map} we can see that the C\,{\sc{iv}}\,1548\,{\AA}\,line is also predominantly redshifted in coronal holes. However, a distinct difference can be found between the C\,{\sc{iv}}\,Dopplergrams in the quiet Sun and coronal holes, i.e.\, there are more pixels with blue shifts in a coronal hole than in an equally-sized quiet-Sun region \citep{Dammasch1999}. Figure\,\ref{fig_qsch_map} shows that Ne\,{\sc{viii}}\,770\,{\AA}\,is blueshifted almost everywhere in a polar coronal hole, which is also distinctly different from the localized blue shifts of Ne\,{\sc{viii}}\,in the quiet Sun. The prevalence of blue shifts has been found in both polar and equatorial coronal holes \citep{Hassler1999,Wilhelm2000,Xia2003,Tu2005a,Aiouaz2005}. These prevalent blue shifts have been observed in not only the Ne\,{\sc{viii}}\,770\,{\AA}\,line, but also in higher-temperature coronal lines such as Fe\,{\sc{x}}\,184.54\,{\AA}\,and Fe\,{\sc{xii}}\,195.12\,{\AA}\,\citep{Tian2010,Tian2011a,Fu2014}. As a result, the spatially averaged Doppler shift in coronal holes is generally blueshifted relative to that in the quiet Sun for spectral lines formed in a wide range of temperatures in the TR and lower corona \citep{Xia2004,Raju2009}. Using SUMER observations, \cite{Xia2004} found that the difference increases with temperature from about $\log\,T= 4.4$ to $\log\,T=6.0$ (Figure\,\ref{fig_qsch_curve}). With EIS observations of more coronal lines, \cite{Tian2010} found that this difference continues to increase with temperature until at least $\log\,T\approx6.3$.

  \begin{figure}    
   \centerline{\includegraphics[width=0.7\textwidth,angle=-90,clip=]{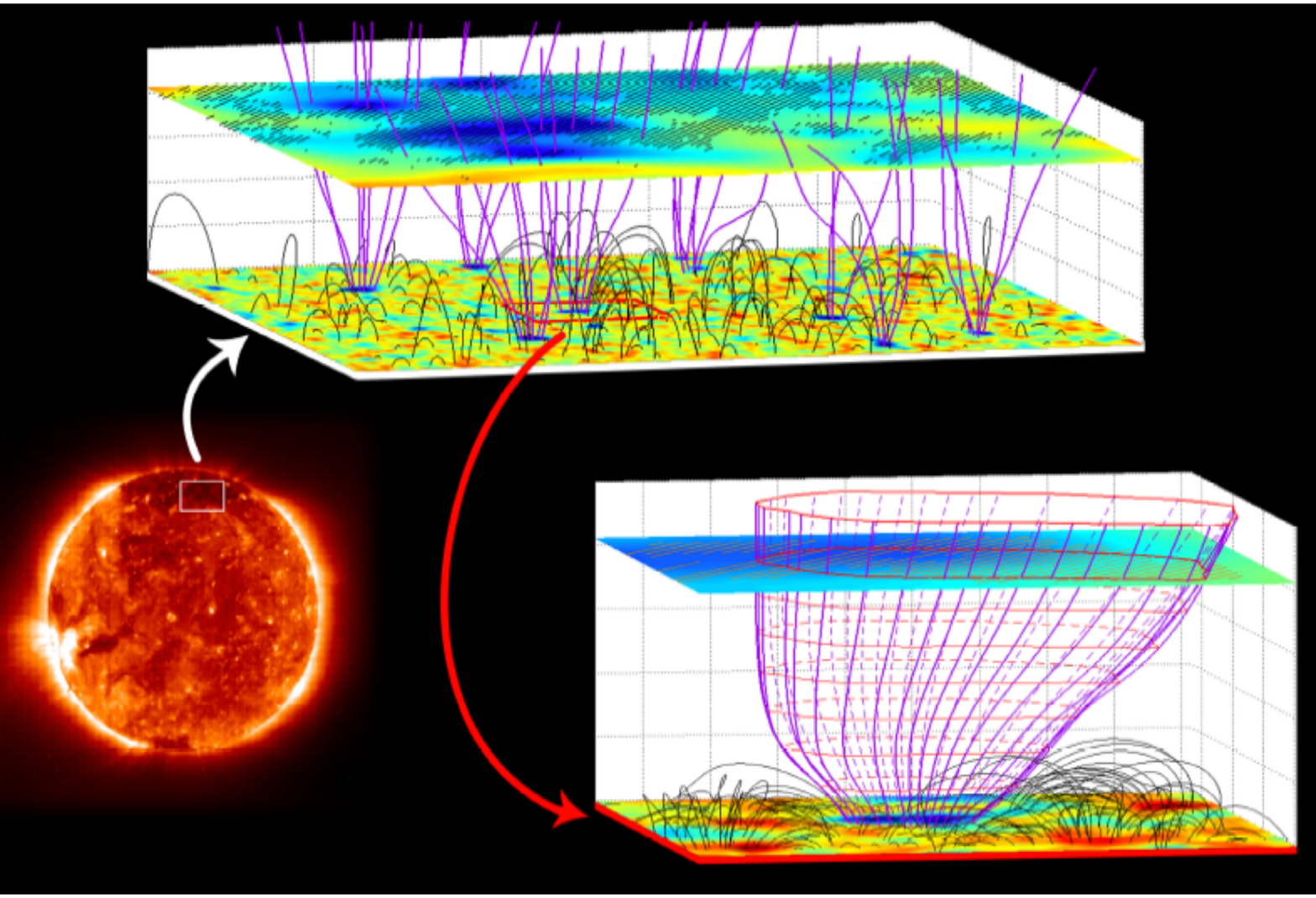}
              }
              \caption{Solar wind origin from funnel-like magnetic structures \citep{Tu2005b}. Lower left: a full-disk coronal image. Top: reconstructed 3D magnetic-field structures in the white rectangular region. The black and purple lines represent closed- and open-field lines, respectively. A photospheric magnetogram (radial component, blue and red colors indicate different polarities) is shown at the bottom. The higher plane shows the magnetogram at 20 Mm, with the hatched region indicating largest blue shifts ($\geqslant$7\,km\,s$^{-1}$) of Ne\,{\sc{viii}}\,770\,{\AA}. Lower right: zoomed-in view of the region indicated by the red rectangle. }
   \label{fig_qsch_wind}
   \end{figure} 
      
The net blue shifts of spectral lines with a formation temperature higher than $\approx$5$\times$10$^{5}$\,K in coronal holes are generally believed to be signatures of the nascent fast solar wind \citep[e.g.\,][]{Hassler1999,Wilhelm2000,Xia2003}. The three-dimensional (3D) coronal magnetic field could be reconstructed from a measured photospheric magnetogram through a potential or force-free field extrapolation, and it can thus be used to correlate with the plasma flow pattern \citep{Wiegelmann2005}. Using a similar method, \cite{Tu2005a} identified many funnel-like magnetic-flux tubes rooted in the chromospheric network of a coronal-hole region, and found that patches of large Ne\,{\sc{viii}}\,blue shift are closely associated with these funnels, suggesting that the nascent fast solar-wind flows outward along these expanding magnetic funnels (Figure\,\ref{fig_qsch_wind}). They calculated the correlation coefficient between maps of the Ne\,{\sc{viii}}\,blue shift and magnetic-field inclination at each height, and they found a maximum correlation at a height of $\approx$20\,Mm above the photosphere. This height could be regarded as a rough estimate of the real formation height of Ne\,{\sc{viii}}\,770\,{\AA}, because the motion of the emitting ions is expected to be largely guided and controlled by the shape of the expanding funnels at the line-emission height in the low-$\beta$ environment.  

It is worth mentioning that apparent jet-like features at coronal temperatures have been frequently reported in coronal holes \citep[e.g.\,][]{Cirtain2007,Ni2020,Shen2021}. Many of these jets are associated with coronal bright points. In addition, \cite{Young2015} found a new type of jet that is visible in the spectroscopic data but shows no clear signature in the imaging data. They are rooted in bright points and have speeds reaching $\approx$100\,km\,s$^{-1}$. These sporadic jets are possibly not the dominant cause of the net blue shifts. 

\subsection{Mass Cycle between the Chromosphere and Corona}
      \label{asymmetry}   

The observed temperature dependence of the Doppler shift is still not well understood. \cite{Tu2005a} and \cite{He2008} interpreted the blue and red shifts in coronal holes as the bidirectional flows generated by magnetic reconnection between the open-field lines in coronal funnels and adjacent low-lying loops. A similar continuous reconnection between legs of large coronal loops (strong network field) and low-lying loops (weak field in internetwork regions) might also exist in the quiet Sun, resulting in bidirectional flows that account for the blue shifts of hotter lines and red shifts of cooler lines \citep{Aiouaz2008}.

  \begin{figure}    
   \centerline{\includegraphics[width=0.99\textwidth,clip=]{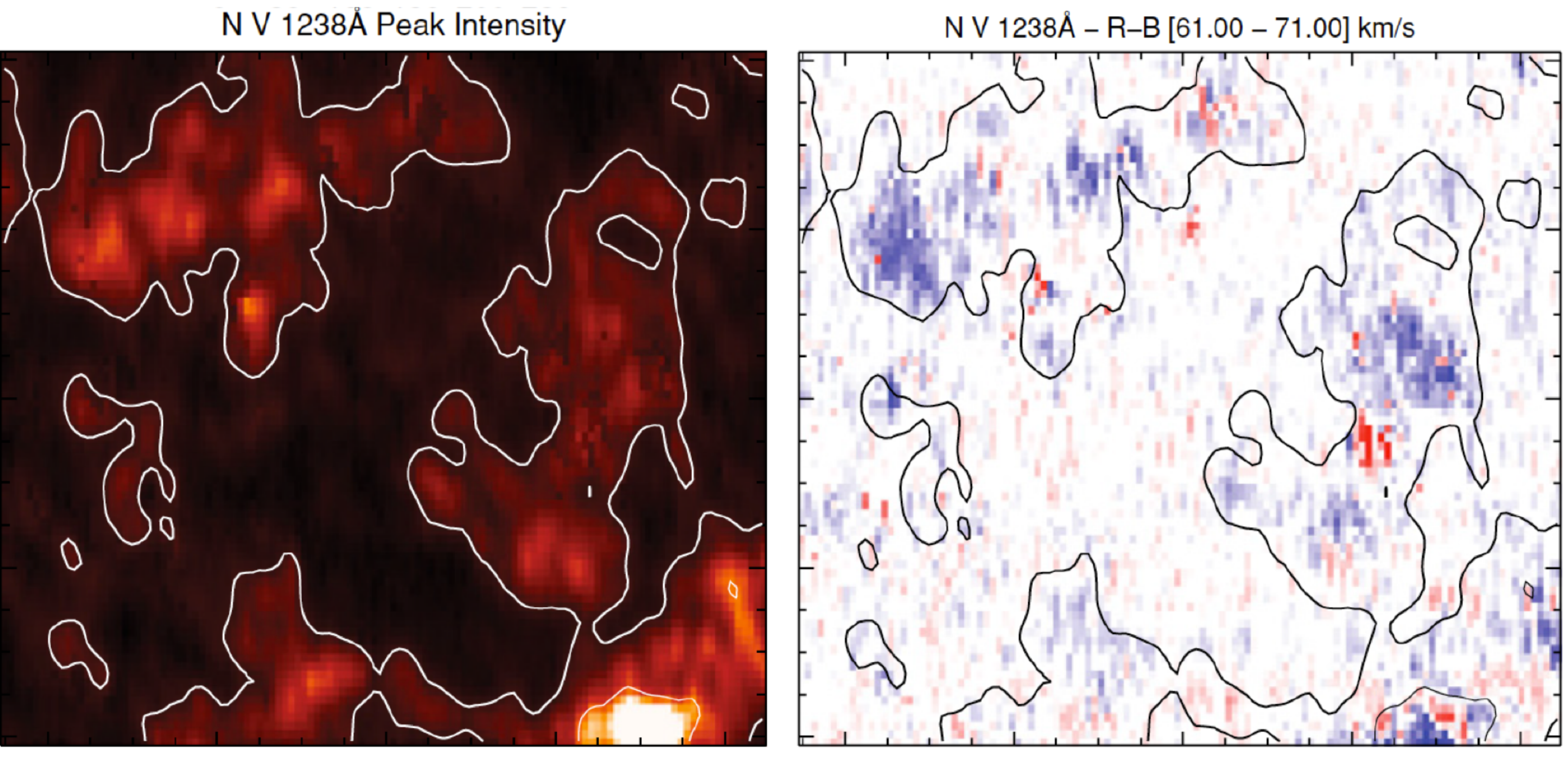}
              }
              \caption{Images of N\,{\sc{v}}\,1238\,{\AA} intensity and profile asymmetry \citep{McIntosh2009a}. Blue and red colors in the asymmetry image indicate enhancements of the 61\,--\,71\,km\,s$^{-1}$ (from line centroid) spectral range at the blue and red wings, respectively. The contours outline regions of strong intensity. The FOV of each image is 90$^{\prime\prime}$$\times$80$^{\prime\prime}$. Reproduced by permission of the AAS. }
   \label{fig_qsch_asy}
   \end{figure}

An alternative interpretation, which has received much more attention in the community, considers the mass-circulation process in the solar atmosphere. Although spectral lines with different formation temperatures show different signs of net Doppler shifts, a careful examination of the TR spectral-line profiles observed by SUMER and IRIS in network regions suggests that the spectral profiles are often enhanced at the blue wing (Figure\,\ref{fig_qsch_asy}). The blue-wing enhancement or blueward asymmetry has been found to occur intermittently, suggesting the presence of episodic weak ($\approx$5\,\% of the line core intensity) high-speed ($\approx$50\,--\,120 km\,s$^{-1}$) upflows \citep{McIntosh2009a,DePontieu2009,ChenY2019}. SUMER observations show that the weak blueward asymmetry can be identified from spectral lines formed in a wide range of temperatures, at least from $\approx$7$\times$10$^{4}$\,K to $\approx$7$\times$10$^{5}$\,K \citep{McIntosh2009a}. These authors interpreted the observed emission of the TR and lower corona as a mixture of emission from rapid injection of episodically heated plasma to the corona and slow cooling of the previously heated plasma. They conjectured that the high-speed plasma ejections could be heating signatures of the high-speed spicules observed in the chromosphere \citep{DePontieu2007}. With simultaneous imaging and spectroscopic observations of the TR, \citet{Tian2014a} discovered prevalent high-speed intermittent jets with a temperature of at least $\approx$10$^{5}$\,K, and they demonstrated that the blueward asymmetry of the Si\,{\sc{iv}}\,1393\,{\AA}\,line is indeed caused by the superposition of these fast TR jets on the TR background. Some of these TR jets have been found to be heating signatures of chromospheric spicules \citep{Pereira2014,Rouppe2015}, whereas others might be classified as the smallest jetlets \citep{Raouafi2014,Panesar2018}. Following the idea that the mass-cycling process is responsible for the observed TR emission, \citet{Wang2013} considered each spectral-line profile as a sum of three components: a high-speed heated upflow generated in the chromosphere, a nearly static background, and a slow cooling downflow. Their detailed investigation suggests that the varying relative contributions of the three components at different temperatures might be responsible for the observed temperature dependence of Doppler shift. The more or less steady behavior of the observed Doppler shifts suggests that such a mass-cycling process should occur continuously.

Numerical simulations have also been performed to understand the temperature dependence of Doppler shift. \cite{Gudiksen2005} developed a 3D magnetohydrodynamic (MHD) model of the solar corona, where the coronal heating is due to the magnetic braiding induced by field line footpoint motions in the photosphere. \cite{Peter2004} and \cite{Peter2006} took this model and synthesized spectra of several TR and coronal lines. They found that this model can reproduce the observed temperature dependence of the TR red shift, implying that the well-known TR red shifts are caused by the flows induced by heating through braiding of magnetic-field lines. However, the model prodicted a net red shift for the Ne\,{\sc{viii}}\,770\,{\AA}\,and Mg\,{\sc{x}}\,625\,{\AA}\,lines that are formed in the upper TR and lower corona, which is inconsistent with SUMER observations but might be due to the relatively low height of the upper boundary.  Using 3D MHD models spanning the upper convection zone to 15 Mm above the photosphere, \cite{Hansteen2010} found that rapid intermittent heating of the plasma from the upper chromosphere to coronal temperatures naturally produces net red shifts for most TR lines and small blue shifts of spectral lines formed above a temperature of $\approx$5$\times$10$^{5}$\,K, roughly consistent with SUMER and EIS observations. These episodic heating events are accompanied by the generation of intermittent high-speed multi-thermal upflows, which may be related to the fast chromospheric spicules \citep{DePontieu2007}, TR network jets \citep{Tian2014a}, and high-speed upflows inferred from the blueward asymmetries of TR lines observed by SUMER \citep{McIntosh2009a}. From high-resolution measurements of the photospheric magnetic field, \citet{Samanta2019} demonstrated that intermittent magnetic reconnection between the strong network field and the newly emerged, weak, small-scale internetwork field is likely responsible for the generation of at least some of these upflowing materials.

\section{Upflows from Active-Region Boundaries} 
      \label{ar}      

The dominant emission structures in the upper solar atmosphere of ARs are large-scale loops that outline the strong magnetic field. Prominent red shifts of spectral lines formed in the TR are commonly observed in these coronal loops, particularly at the loop legs \citep[e.g.\,][]{Winebarger2002,Marsch2004,Dammasch2008,Marsch2008,DelZanna2008}. Different from the TR red shifts observed in the network lanes of quiet-Sun regions and coronal holes, these red shifts are also present in coronal lines with a formation temperature as high as $\approx$10$^{6.2}$\,K \citep{DelZanna2008}. The magnitude of these red shifts could reach $\approx$30\,km\,s$^{-1}$, two to three times higher than in the quiet Sun and coronal holes. Despite these differences, the origin of these red shifts is likely to be similar to that in the quiet Sun and coronal holes \citep[e.g.\,][]{Dammasch2008}.

\begin{figure}    
   \centerline{\includegraphics[width=0.99\textwidth,clip=]{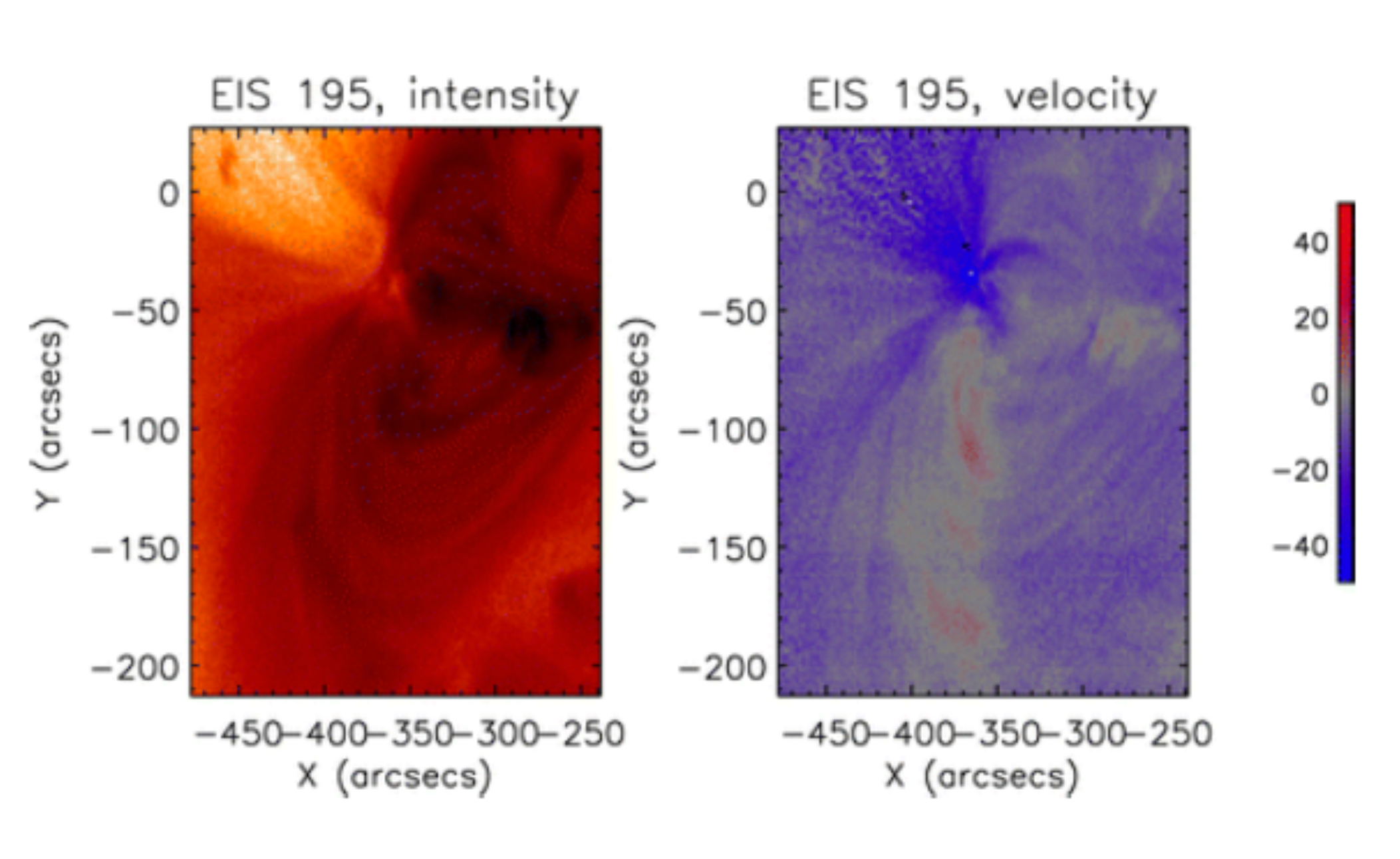}
              }
              \caption{Intensity image (shown with a reversed color table) and Dopplergram of the Fe\,{\sc{xii}}\,195.12\,{\AA}\,line obtained during a raster scan of \textit{Hinode}/EIS \citep{Harra2008}. Reproduced by permission of the AAS. }
   \label{fig_ar_blueshift}
   \end{figure}

At the boundaries of many ARs, the coronal emission is normally weaker than in the AR cores. With SOHO/SUMER observations, \cite{Marsch2004} found signatures of upflows in the Ne\,{\sc{viii}}\,770\,{\AA}\,line, which is formed in the upper TR or lower corona, at the boundaries of ARs. With \textit{Hinode}/EIS observations, coronal spectral lines such as Fe\,{\sc{xii}}\,195.12\,{\AA}\,and Fe\,{\sc{xiii}}\,202.04\,{\AA}\,are often found to be blueshifted \citep[e.g.\,][]{Harra2008} (Figure\,\ref{fig_ar_blueshift}). This is one of the major discoveries of EIS. Early results about these blueshifts have been summarized by \citet{Harra2012a}, and there has also been a short summary of studies connecting the upflows to the solar wind using EIS observations \citep{Hinode2019}. In this section we provide an updated and more extended review of past investigations on these upflows. 
   
\subsection{Velocity Measurements} 
  \label{velocity}
  
These blue shifts appear to be quasi-steady, i.e.\, they generally last for at least a few days \citep[e.g.\,][]{Baker2009,Bryans2010,Tian2012a}. A single Gaussian fit (SGF) to the observed coronal line profiles often leads to a blue shift of $\approx$10\,--\,50\,km\,s$^{-1}$ \citep[e.g.\,][]{Harra2008,Marsch2008,Doschek2008,Srivastava2014}. These blue shifts are associated with enhanced line broadenings (or nonthermal velocities), and there is a positive correlation between them \citep[e.g.\,][]{Doschek2008}. From magnetic-field extrapolations, these upflows appear to be guided by the legs of large-scale magnetic loops or open magnetic-field structures at AR boundaries \citep[e.g.\,][]{Marsch2004,Harra2008,Marsch2008,Baker2009}. These structures expand rapidly with height, often appearing as fan-like structures. The upflows are normally more obvious towards the footpoints of the fans, but are not always associated with them \citep{Warren2011}. They are, however, sometimes very prominent in low-emission regions at the peripheries of AR cores or between the cores and fans \citep[e.g.\,][]{McIntosh2012,Scott2013}. 

\citet{Demoulin2013} and \citet{Baker2017} tracked several ARs as they crossed the solar disk. By applying a simple model of stationary upflow to the observed ARs, they found that the long-term evolution of the persistent upflows is consistent with the scenario of steady upflows projected onto the line of sight (LOS). They found no obvious dependence of the de-projected upflow velocity on the AR age. They were also able to determine the inclination angles of the magnetic-field lines that guide the upflows. \citet{Baker2017} found that the inclination angles are typically in the range of [0$^{\circ}$,40$^{\circ}$] and  [-30$^{\circ}$,30$^{\circ}$] relative to to the local vertical direction for the trailing and leading polarities, respectively. These angles are roughly consistent with those inferred from magnetic-field extrapolations. The upflow velocities show no obvious difference between the trailing and leading polarities and do not depend on the underlying photospheric magnetic-field strength. They also found that transient coronal events such as CMEs, jets, and flares could lead to a variation of the upflow velocities on different time scales. However, these transient changes normally do not affect the quasi-steady behavior of the upflows.    

These upflows first appear during the phase of flux emergence and generally persist during the whole lifetimes of ARs. \citet{Harra2010} performed a detailed investigation of an emerging AR, and they witnessed the formation of these upflows. As the AR expands, they observed a ring of blue shifts in coronal spectral lines around the edge of the emerging AR. The upflows clearly intensify as more magnetic flux emerges. In a subsequent study, \citet{Harra2017} tracked the evolution of a decaying AR during three solar rotations, and they found a larger area occupied by upflows and slightly higher upflow speeds as the AR evolves. \citet{Zangrilli2016} extended investigations of AR upflows on the solar disk to the off-limb corona utilizing observations from the \textit{Ultraviolet Coronagraph Spectrometer} \citep[UVCS:][]{Kohl1995} onboard SOHO. Their findings demonstated that upflows persist over the lifetime of an AR spanning over four solar rotations. 

  \begin{figure}    
   \centerline{\includegraphics[width=0.99\textwidth,clip=]{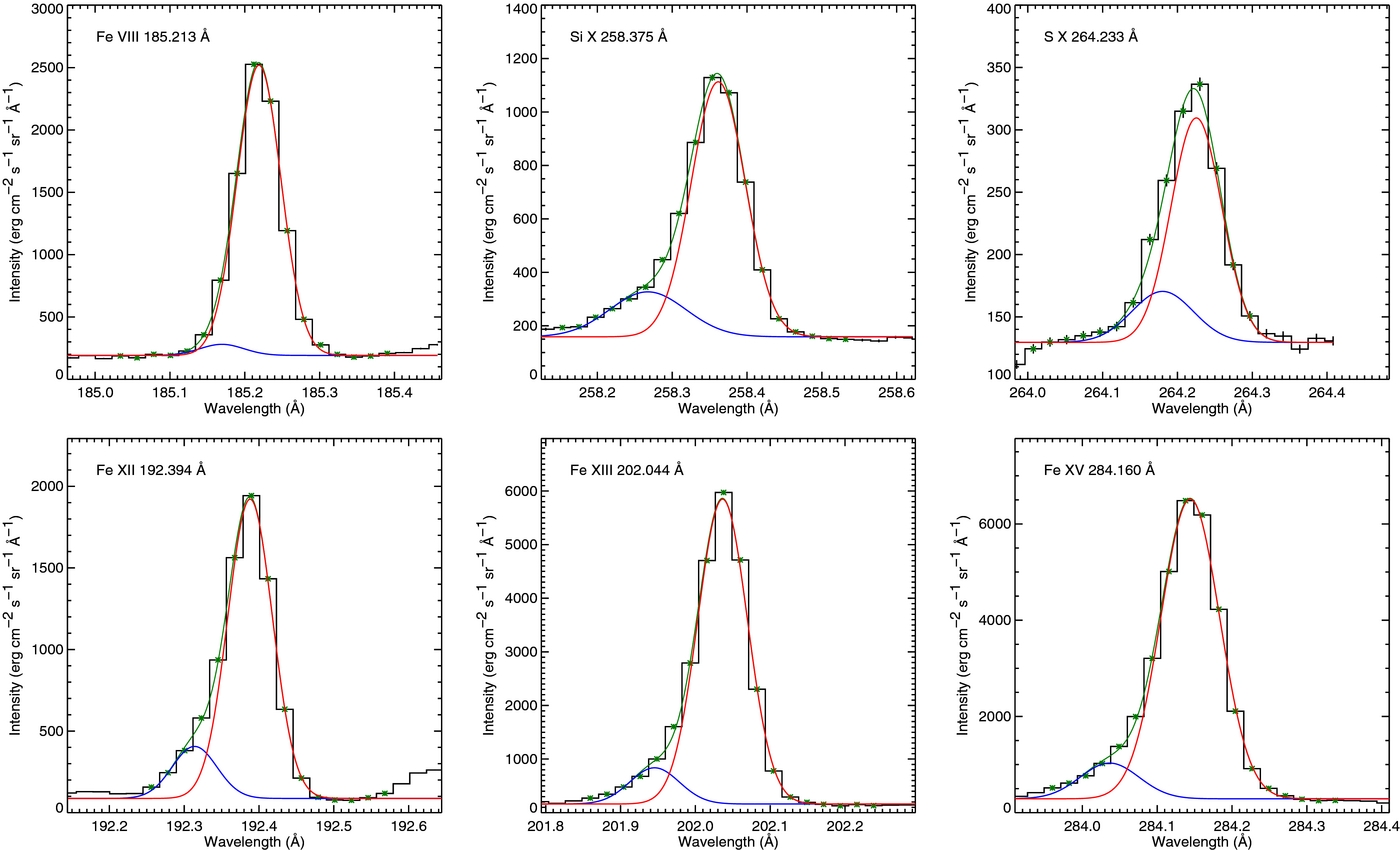}
              }
              \caption{Double-Gaussian fits to several line profiles observed at an AR boundary \citep{Brooks2012}. The histograms represent the observed line profiles. In each panel, the primary and secondary components are shown as the red and blue curves, respectively. The green curve represents the total of the two components. Reproduced by permission of the AAS. }
   \label{fig_ar_profile}
   \end{figure}
   
Through an examination of the coronal line profiles at AR boundaries, \citet{Hara2008} found that the profiles are generally asymmetric, with a weak yet noticeable enhancement at the blue wings. These blueward asymmetries suggest the presence of unresolved high-speed upflows. Later, detailed analyses of these line profiles suggest that the coronal emission consists of at least two components: a primary component accounting for the nearly stationary background emission and a secondary component associated with high-speed upflows \citep[e.g.\,][]{Tian2011b,Brooks2012}. Figure\,\ref{fig_ar_profile} shows examples of such line profiles, which also indicate a temperature dependence of the high-speed component. Several methods have been introduced to separate the two components and characterize the weak high-speed component, including the technique of double-Gaussian fit \citep{Peter2010,Bryans2010,Brooks2012,Doschek2012,Kitagawa2015}, red--blue (RB) asymmetry analysis \citep{DePontieu2009,DePontieu2010,Martinez-Sykora2011,Tian2011b,Tian2011c} and RB-guided double-Gaussian fit \citep{DePontieu2010,Tian2011b}. Application of these methods to the observed coronal spectral-line profiles at AR boundaries mostly yielded speeds on the order of $\approx$100\,km\,s$^{-1}$, often in the range of 50\,--\,150\,km\,s$^{-1}$, for these high-speed upflows. Occasionally the velocity may reach $\approx$200 km\,s$^{-1}$. The primary component is often also blueshifted, but by only $\approx$10 km\,s$^{-1}$. The intensity ratio of the two components is mostly less than 15\,\%, although at some locations it could reach more than 30\,\%. The widths of the two components are often comparable, which has also been confirmed through a comparison with analysis of artificial profiles \citep{Tian2011b}. Since an observed spectral-line profile is the superposition of the two components, an enhanced line broadening and a moderate blue shift ($\approx$10\,--\,50\,km\,s$^{-1}$) are expected if a SGF is applied. This superposition also explains the observed positive correlation of the blueward asymmetry with the blue shift and line broadening determined from a SGF \citep{Tian2011b}. It is worth mentioning that a blue-wing enhancement could in principle result from the superposition of multiple (more than two) emission components, each slightly Doppler-shifted with respect to each other \citep{Doschek2012}. This scenario might explain some of the AR upflows but is not very likely for all of them, because at some locations the blue-wing enhancement is so strong and far from the line core that a multiple-component scenario is really not consistent with the observed line profiles. In addition, speeds of the counterparts of these upflows in coronal image sequences (see Section\,\ref{waves}) can often be unambiguouly measured and they do not show a continuous distribution. 

  \begin{figure}    
   \centerline{\includegraphics[width=0.99\textwidth,clip=]{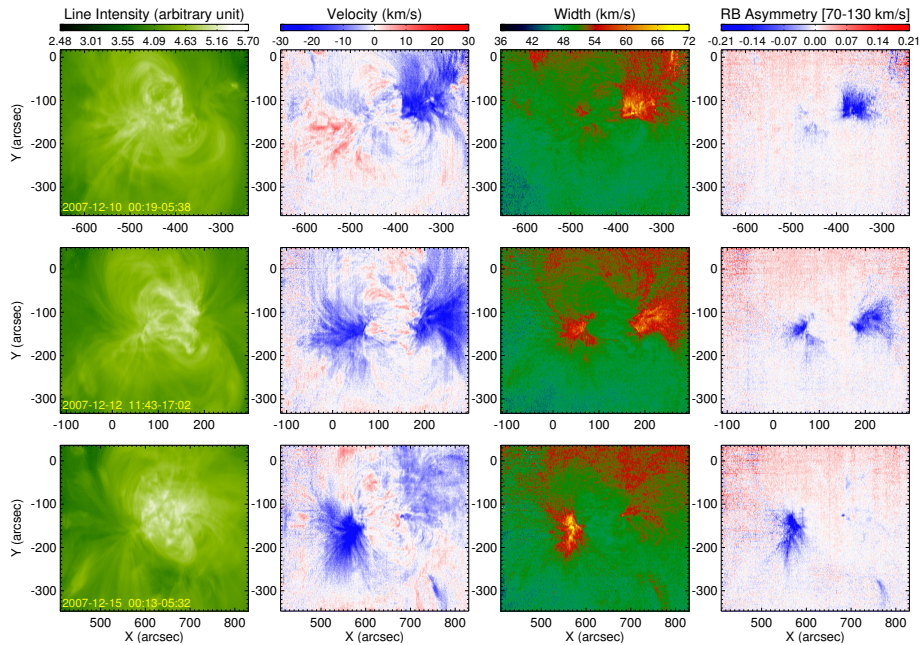}
              }
              \caption{Maps of the SGF parameters and profile asymmetry in the velocity interval of 70\,--\,130\,km\,s$^{-1}$ \citep{Tian2012a}. The first, second, and third rows correspond to observations of the same AR when it is located near the east limb, disk center, and west limb, respectively. Reproduced by permission of the AAS. }
   \label{fig_ar_clv}
   \end{figure}

Although the blueward asymmetries are weak, they are definitely not caused by random noise or blends by other spectral lines. \citet{Tian2012a} have demonstrated this by tracking an AR as it rotates from the east limb to the disk center and then to the west limb. From Figure\,\ref{fig_ar_clv} we see a clear center-to-limb variation of the line parameters. Here the RB asymmetry is defined as the difference of the two wing intensities integrated over the velocity interval of 70\,--\,130\,km\,s$^{-1}$ divided by the peak intensity. When the AR is located near the disk center on 12 December 2007, we see prominent blue shifts, large line broadenings, and blueward asymmetries at both boundaries. As the AR rotates to the west limb on 15 December 2007, the profile asymmetries disappear at the western boundary. This phenomenon can be understood if we consider the scenario of a high-speed field-aligned upflow superimposed on a nearly static coronal background. When the AR is close to the west limb, the magnetic-field lines at the western boundary are roughly perpendicular to the LOS so that the projected speed of the field-aligned flow on the LOS direction is very small. This leads to a greatly reduced blueward asymmetry as well as a reduced Doppler shift and line broadening. Similarly, when the AR is located near the east limb on 10 December 2007, the blue shift, line broadening, and blueward asymmetry are all reduced at the eastern boundary because the field lines are roughly perpendicular to the LOS there. 

  \begin{figure}    
   \centerline{\includegraphics[width=0.72\textwidth,angle=-90,clip=]{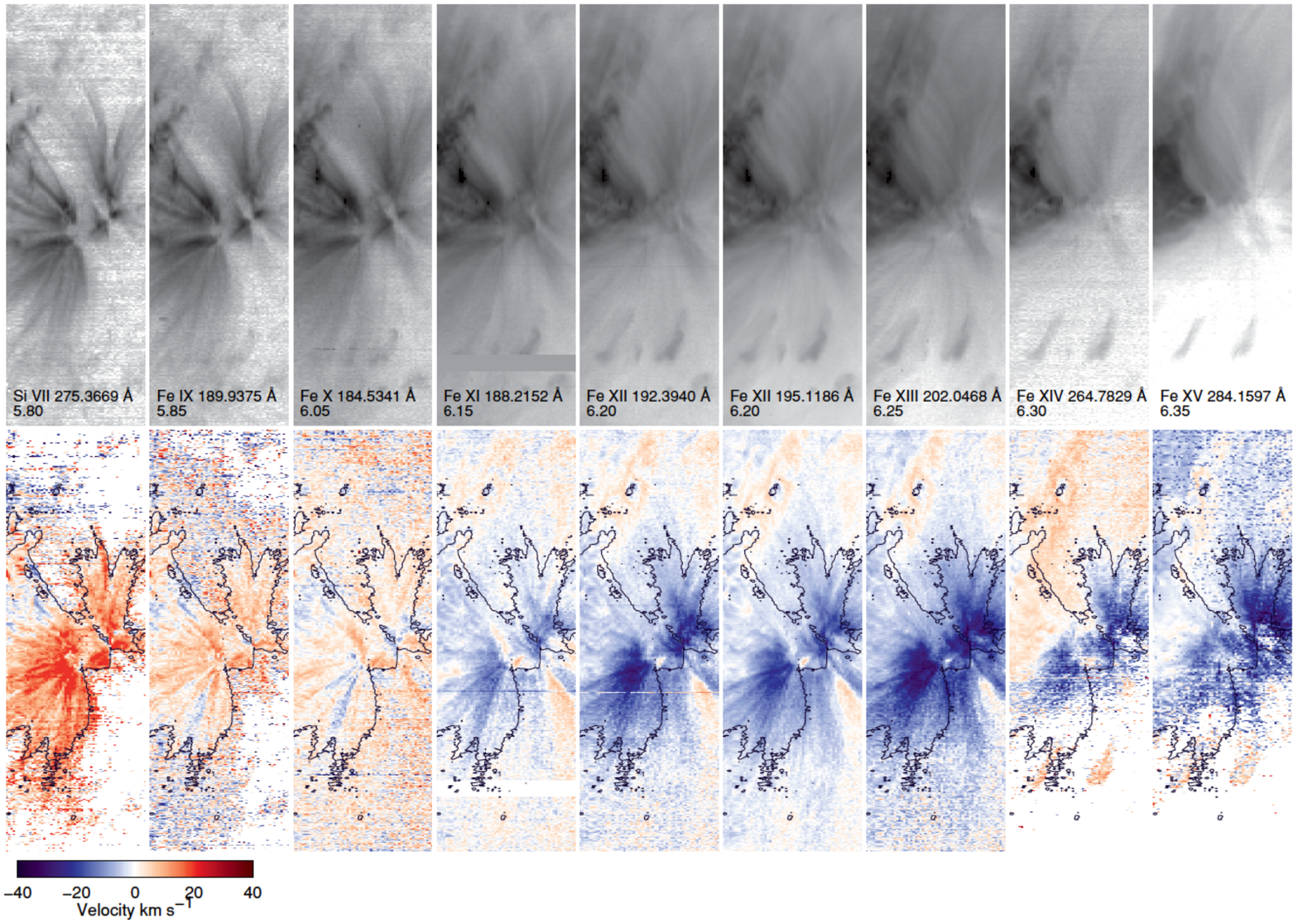}
              }
              \caption{Intensity images (shown with a reversed color table) and Dopplergrams of several emission lines at an AR boundary \citep{Warren2011}. Contours of the Si\,{\sc{vii}}\,275.35\,{\AA}\,intensity is overlaid on the Dopplergrams. The FOV of each image is about 180$^{\prime\prime}$$\times$512$^{\prime\prime}$. Reproduced by permission of the AAS. }
   \label{fig_ar_red2blue}
   \end{figure}
  
There is a clear dependence of the SGF velocity on the line-formation temperature at AR boundaries. \citet{DelZanna2008} found that the SGF blue shift increases from only a few km\,s$^{-1}$ at 10$^{5.8}$\,K to $\approx$30\,km\,s$^{-1}$ at 10$^{6.3}$\,K at an AR boundary. \citet{Tripathi2009} analyzed the EIS data at another AR boundary. They found an average red shift of a few km\,s$^{-1}$ for the Si\,{\sc{vii}}\,275.35\,{\AA}\,line formed at $\log\,T$\,$\approx$\,5.8, and an increase of the average blue shift from $\approx$2\,km\,s$^{-1}$ at 10$^{6.0}$\,K to $\approx$10\,km\,s$^{-1}$ at 10$^{6.3}$\,K. Actually, at AR boundaries both downflows and upflows could exist in the TR line Si\,{\sc{vii}}\,275.35\,{\AA}. \citet{McIntosh2012} found that this line is redshifted at the fan-like structure of an AR and blueshifted in a dark region between the fan and the AR core. If we only consider fan-like structures, a clear trend from red shifts to blue shifts is normally observed in the temperature range of $\log\,T$\,=\,5.8 to 6.3 (Figure\,\ref{fig_ar_red2blue}). The transition from red shifts to blue shifts occurs at a temperature of $\log\,T$\,$\approx$\,6.0, meaning that spectral lines formed at typical coronal temperatures and TR temperatures are predominantly blueshifted and redshifted, respectively. We note, however, that there are cases where the morphology of the fans and upflows is different, and there is no fan emission at low temperatures beneath the upflows \citep{Warren2011}.

  \begin{figure}    
   \centerline{\includegraphics[width=0.99\textwidth,clip=]{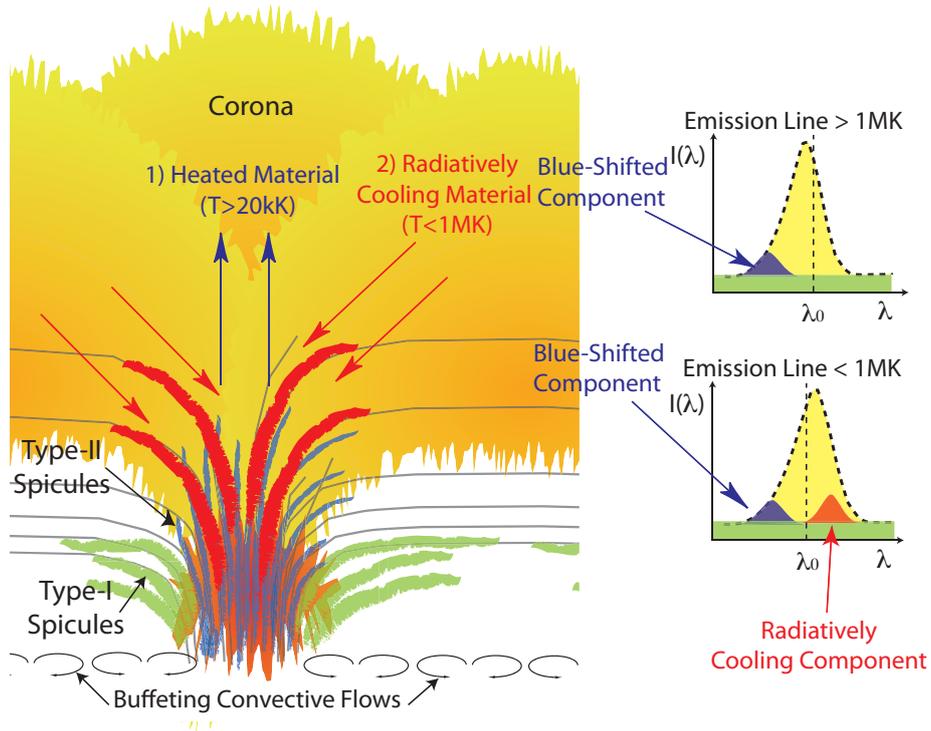}
              }
              \caption{A cartoon showing the mass cycling between the chromosphere and corona \citep{McIntosh2012}. Reproduced by permission of the AAS. }
   \label{fig_ar_3components}
   \end{figure}
   
By examining the EIS line-profile asymmetries and slot images, \citet{UgarteUrra2011} have attempted to understand the different Doppler shift patterns of coronal lines and TR lines. They identified clear outward-propagating disturbances with speeds of 40\,--\,130\,km\,s$^{-1}$ from the Fe\,{\sc{xii}}\,195.12\,{\AA}\,image sequence, consistent with the quasi-periodic blue-wing enhancements in the Fe\,{\sc{xii}}\,line profiles. From the Si\,{\sc{vii}}\,275.35\,{\AA}\,slot images, they could not identify any obvious signature of outward-propagating features. Instead, they identified downward-propagating features with speeds of 15\,--\,20\,km\,s$^{-1}$, which are comparable to the magnitude of red shifts at the locations of fans. From imaging observations of AIA, \citet{Kamio2011} and \citet{McIntosh2012} have also identified quasi-periodic fast ($\approx$100\,km\,s$^{-1}$) outward-propagating disturbances in the hot 193\,{\AA}\,channel and sporadic slow ($\approx$15\,km\,s$^{-1}$) downflows in cool channels such as 131\,{\AA}. \citet{McIntosh2012} also found a mixture of upflows and downflows from the 171\,{\AA}\,images, indicating that the dominant flow changes its direction around a temperature of $\approx$1\,MK.
    
These observations imply a mass-cycling process between the chromosphere and corona (Figure\,\ref{fig_ar_3components}). As \citet{McIntosh2012} suggested, local chromospheric materials may be impulsively heated, producing high-speed upflows with typical TR and coronal temperatures. As these upflows experience significant radiative cooling, the previously heated and injected plasma will slowly drain to the chromosphere. These downflows may have a typical temperature of the TR plasma. EIS observations have shown that TR lines such as Si\,{\sc{vii}}\,275.35\,{\AA}\,also reveal blueward asymmetries \citep{Tian2011b,McIntosh2012}. However, the velocity of the secondary component inferred from Si\,{\sc{vii}} line profiles appears to be smaller than that inferred from the simultaneously observed coronal line profiles, which is likely due to the fact that TR line profiles are complicated by downflows. Unlike the two-component coronal line profiles, a TR line profile likely consists of three components: a fast upflow component generated through chromospheric heating, a slow cooling downflow component, and a TR background component. The downflows have a much lower speed and possibly stronger emission compared to the upflows, which would result in net red shifts of the TR lines when a SGF is performed. So similar to the quiet Sun and coronal holes, the temperature dependence of the SGF Doppler shift at AR boundaries is probably caused by the relative contributions of the different components at different temperatures. A similar scenario of mass cycling or circulation has also been proposed by \citet{Marsch2008} and \citet{Young2012}.

Sometimes the red-shifted downflows on the fan loops appear to be spatially uncorrelated with the blue-shifted upflows \citep[as in one of the examples in][]{Warren2011}. In these cases the mass-circulation picture might still be relevant if the upflows are connected to a distant AR \citep[as in][]{Boutry2012}, i.e. the red shifts on the fans of the distant AR might be signatures of the circulation. In such cases the morphology of the fan loops and upflows is similar because of the general topology of the magnetic field at the active-region boundary \citep{Baker2009}. 

\subsection{Waves or Flows} 
  \label{waves}

  \begin{figure}    
   \centerline{\includegraphics[width=0.99\textwidth,clip=]{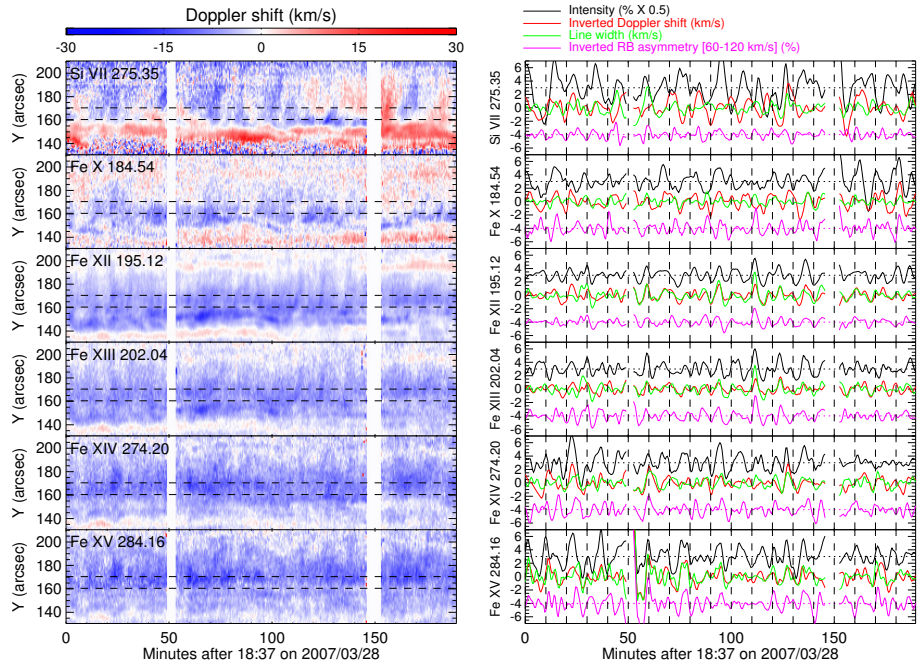}
              }
              \caption{Quasi-periodic variation of the AR upflows \citep{Tian2012b}. Left:  Temporal evolution of the Doppler shift of several emission lines. Right: Curves with different colors represent the temporal changes of the line intensity, Doppler shift, line width, and profile asymmetry (RB asymmetry) averaged within the region between the two dashed lines in each of the left panels. For a clearer illustration, the intensity and asymmetry curves are offset by 3 and -4, respectively. Reproduced by permission of the AAS. }
   \label{fig_ar_timeseries}
   \end{figure}

From EUV and X-ray imaging observations, we often find quasi-periodic upward propagating disturbances (PDs) along the fan-like structures at AR boundaries \citep[e.g.\,][]{deMoortel2000,Sakao2007,Yuan2012}. These PDs have a propagation speed of $\approx$50\,--\,200\,km\,s$^{-1}$, and they often recur at a time scale of 3\,--\,15 minutes. These PDs were widely interpreted as slow-mode magnetoacoustic waves propagating along the fans, mainly because the speed is comparable to the coronal sound speed and there is a correlation between the intensity and velocity perturbations \citep{deMoortel2009,Wang2009a,Wang2009b,Nishizuka2011}. 

The aforementioned blue shifts and these PDs are often found at the same locations, indicating that they are closely related to each other. Based on this coincidence, the PDs have been suggested to be plasma upflows \citep{Sakao2007,Harra2008}. However, there appear to be some distinct differences between them. First, around the year 2008, it was unclear why the blue shifts  are only $\approx$10\,--\,50\,km\,s$^{-1}$, which are significantly lower than the speeds of PDs. A projection effect has been proposed to explain the apparent discrepancy between the LOS velocities of the blueshifts and the plane-of-sky (POS) velocities of the PDs. However, the blue shifts seldom exceed 50\,km\,s$^{-1}$, regardless of AR location on the solar disk. Second, the PDs are clearly quasi-periodic. However, it was unclear whether the blue shifts showed a quasi-periodic variation around 2008.

These discrepancies disappear if we compare the PDs with the high-speed secondary components of coronal line profiles. From a coordinated observation with XRT and EIS onboard \textit{Hinode}, \citet{Tian2011c} found a clear correspondence between the fluctuations of the X-ray intensity and the blueward asymmetries of coronal spectral lines, suggesting that the PDs observed in XRT images are closely related to the secondary components. Through joint observations of SDO/AIA and \textit{Hinode}/EIS, \citet{Tian2011b} demonstrated that the velocity distributions and relative intensities are both remarkably similar for the simultaneously observed PDs and secondary components. In addition, with sit-and-stare observations of EIS, \citet{DePontieu2010} and \citet{Tian2011c} noticed that the blue shifts and blueward asymmetries often show quasi-periodic variations. A following statistical study by \citet{Tian2012b} demonstrated that this behavior is quite common, and that the recurring time scale is similar to that of PDs. So it appears that the PDs are related to the high-speed secondary components.

Detailed analyses suggest that the blueward asymmetry as well as the intensity, blue shift and line width derived from a SGF all reveal correlated quasi-periodic variations. Figure\,\ref{fig_ar_timeseries} shows an example. The quasi-periodic enhancement of blue shift is clearly seen from the space--time diagrams. For the spectral lines formed at typical coronal temperatures of $\log\,T$\,=\,6.0\,--\,6.3 (Fe\,{\sc{x}}\,184.54\,{\AA}, Fe\,{\sc{xii}}\,195.12\,{\AA}, Fe\,{\sc{xiii}}\,202.04\,{\AA}, Fe\,{\sc{xiv}}\,274.20\,{\AA}, and Fe\,{\sc{xv}}\,284.16\,{\AA}), we can see correlated changes of all these line parameters. Such correlated changes can be explained by a scenario of recurring or quasi-periodically enhanced upflows \citep{DePontieu2010,Tian2011c,Tian2012b}. On occasions when there is only the nearly stationary background-emission component and no upflow, the Doppler shift is zero and the spectral profile is symmetric. While when a high-speed upflow component coexists with the background component, a SGF to the total emission profile will lead to an increase in the line intensity and line width. In the meantime, the line profile reveals an enhancement in the blue wing and a SGF gives a small blue shift. This scenario naturally explains the in-phase variations of all line parameters, which appears to favor the interpretation of intermittent plasma jets for the PDs. If this new interpretation is correct, previous diagnostics of the coronal plasma based on the wave interpretation of PDs would be significantly impacted. 

However, there is also observational evidence supporting the slow-wave interpretation. For instance, \citet{Verwichte2010} considered a scenario of propagating slow waves with a quasi-static background plasma component in the LOS. They were able to reproduce a blueward asymmetry in the line profiles, and a correlation of the blueward asymmetry with the line intensity and blue shift. They concluded that the slow-wave interpretation is still valid for the PDs. However,  their model predicted a double frequency (or half period) in the line width that is not seen in EIS observations. There are also a couple of studies reporting a temperature-dependent PD velocity \citep{KrishnaPrasad2012,Uritsky2013}, which is a characteristic of slow-mode magneto-acoustic waves. However, several other studies did not reveal any obvious temperature dependence \citep{Tian2011b,Sharma2020}. A statistical study by \citet{Kiddie2012} showed that the temperature dependency appears to be clear only for PDs rooted in sunspots. 

  \begin{figure}    
   \centerline{\includegraphics[width=0.78\textwidth,angle=-90,clip=]{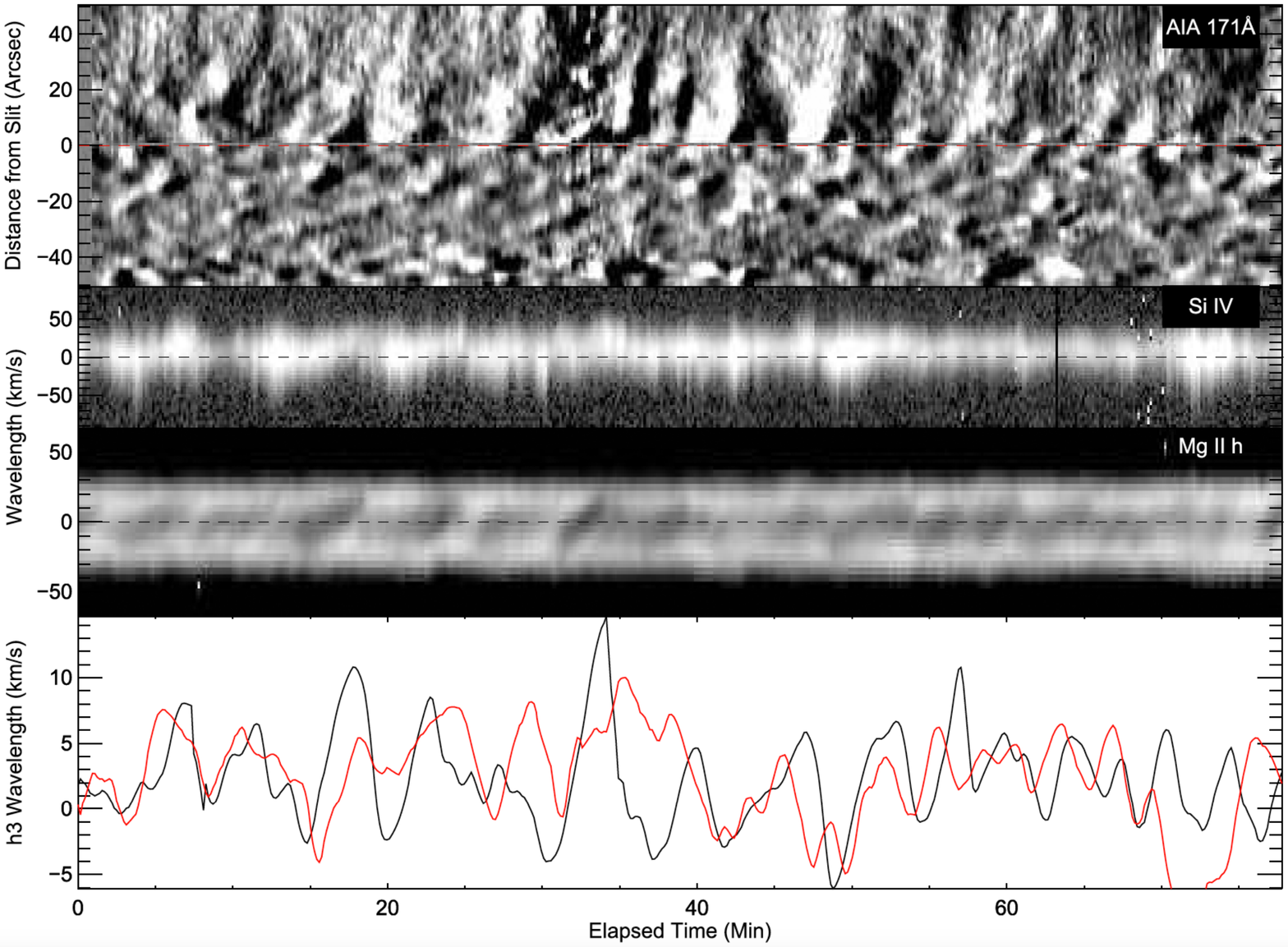}
              }
              \caption{Contribution of PDs from both waves and flows \citep{Bryans2016}. Top panel: a space--time diagram of the AIA 171\,{\AA}\,running difference for a virtual slit along the leg of a fan loop. The red horizontal line indicates the IRIS slit location. Second and third panels: temporal evolution of the Si\,{\sc{iv}}\,1403\,{\AA}\,and Mg\,{\sc{ii}}\,h\,1403\,{\AA}\,spectral line profiles. The black horizontal lines indicate the rest wavelengths of the two lines. Bottom panel: temporal evolution of the Mg\,{\sc{ii}}\,h3 Doppler shift (black) and the negative of the 171\,{\AA}\,intensity (normalized, red). Reproduced by permission of the AAS. }
   \label{fig_ar_waveflow}
   \end{figure}

After several years of debate, there is now a consensus that both waves and flows exist at AR boundaries. For example, recent IRIS and AIA observations have revealed signs of magneto-acoustic shock waves and jet-like flows, both of which appear to show a correspondence to the PDs \citep{Bryans2016}. As shown in Figure\,\ref{fig_ar_waveflow}, the quasi-periodic sawtooth-like patterns in the Mg\,{\sc{ii}}\,h\,1403\,{\AA}\,spectral profiles \citep[evidence of magnetoacoustic shock waves, e.g.][]{Rouppe2003,Tian2014b,Skogsrud2016}, the Si\,{\sc{iv}}\,blue shifts (signs of jet-like flows) and the PDs in 171\,{\AA}\,appear to be correlated. Thus both waves and flows may contribute to the signals of PDs. From EIS observations, the enhanced line width and blueward asymmetry are most obvious at the lower parts of fan structures, and they become absent at higher parts. This observational fact might be explained by a LOS projection effect or lower signals at larger heights. However, it could also be explained as due to the dominance of slow-mode waves away from the footpoint regions. Indeed, \citet{Nishizuka2011} found an in-phase variation of the coronal-line intensity and Doppler shift at larger heights, which was interpreted as an evidence of slow-mode waves. They also claimed that at lower heights fast upflows obviously exist and may be driven by heating events around the footpoints of loops.

Attempts have been made to understand the coexistence of waves and flows. For instance, \citet{Ofman2012} performed 3D MHD modeling of hot ($\approx$6\,MK) AR loops, and found that impulsively and periodically driven upflows can excite undamped slow-mode waves that propagate along the magnetic loops. Their simulations also generated slow-mode shock-like wave trains when the driving pulses have a large amplitude. In a subsequent numerical investigation, \citet{WangT2013} expanded this model to warm loops ($\approx$1\,MK) and found a similar result. They concluded that at lower heights both the flows and waves contribute to the PDs, while at larger heights the PDs are likely dominated by slow waves, as the flows decelerate during the upward propagation. However, these simulations assumed an isothermal plasma, the effect of which needs to be investigated in the future. In addition, the speeds of the upflow pulses in these simulations appear to be much smaller than the speeds of the secondary components inferred from EIS observations.

\subsection{Generation Mechanisms} 
  \label{generation}
   \begin{figure}    
   \centerline{\includegraphics[width=0.98\textwidth,angle=0,clip=]{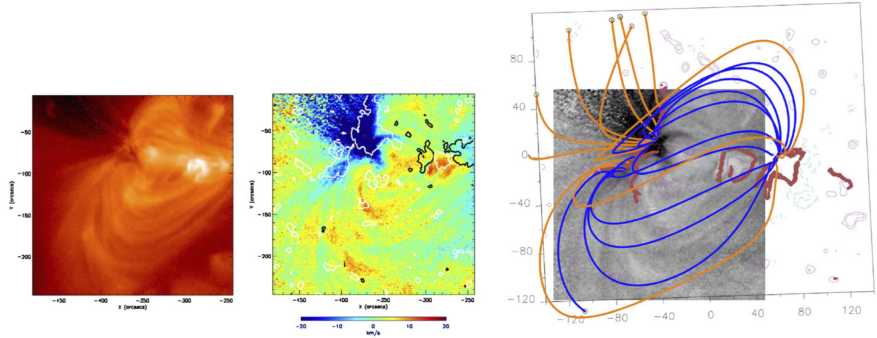}
              }
              \caption{\emph{Hinode}/EIS Fe {\sc xii}\,195\,{\AA} intensity (left) and Doppler velocity (middle) maps and photospheric traces of QSLs in thick red lines overplotted on a grayscale Dopplergram (right).  MDI magnetic field isocontours of $\pm$50\,G are shown in white/black (middle).  Field lines with circles leave the computational box and are considered to be open field.  Strong upflows (blue patches in the middle panel, or dark patches in the right panel) over the positive polarity are along open-field lines on the eastern edge of AR 10942 \citep{Baker2009}. The EIS images correspond to a FOV of about 250$^{\prime\prime}$$\times$250$^{\prime\prime}$. Reproduced by permission of the AAS. }
   \label{fig_qsls}
   \end{figure}
 
\citet{Baker2009} investigated the magnetic-field topology of an AR using a linear force-free field extrapolation method. From the reconstructed 3D coronal magnetic field, they identified quasi-separatrix layers (QSLs), which are locations of strong gradients of magnetic connectivity. The strongest upflows appear to be located in the vicinity of QSL sections over areas of strong field. Based on this finding, the authors suggested that the upflows are driven by magnetic reconnection at QSLs between closed field lines at AR cores and open-field lines or large-scale externally connected loops from AR boundaries. \citet{Edwards2016} analyzed EIS data of seven ARs and found that the upflows generally do not  correspond to high-reaching loops or open-field structures predicted by the global potential-field source-surface model, but they also found that the upflows often coincide with the footprints of separatrix surfaces that are associated with coronal null points. A good agreement between upflow regions and QSLs, either temporally or spatially or both, has also been found by \citet{Scott2013}, \citet{Demoulin2013}, \citet{Mandrini2015}, and \citet{Baker2017}. Similarly, \citet{Demoulin2013} proposed that the upflows are driven by the upward pressure gradient after magnetic reconnection between the high-pressure AR loops and neighboring low-pressure loops. This scenario naturally explains the observational fact that stronger upflows are located closer to the AR cores. The steadiness of the upflows could be understood if we consider the scenario of  successive reconnections, which leads to a superposition and thus averaging of flows with different velocities. 

\citet{DelZanna2011} studied two ARs and found a clear association of the upflows with metric radio noise storms and large-scale open separatrix field lines. Based on this connection, they proposed that the upflows are driven by interchange reconnection between magnetic loops in AR cores and adjacent open-field structures. In this scenario, continuous AR expansion leads to successive reconnection at coronal null points, which results in a strong pressure gradient that drives the temperature-dependent plasma upflows. \citet{Bradshaw2011} numerically simulated this scenario and found the development of a rarefaction wave in the post-reconnection region. Their forward calculation yielded a $\approx$10\,--\,50\,km\,s$^{-1}$ blueshift and clear temperature dependence of the velocity magnitude, consistent with observational results based on a SGF \citep[e.g.\,][]{DelZanna2008}. On the other hand, a 3D data driven simulation of a similar mechanism by  \citet{Galsgaard2015} failed to reproduce systematic upflows seen in EIS observations \citep{Vanninathan2015}. Instead, this model generated magneto-acoustic waves.  

Besides magnetic reconnection, AR expansion has also been considered as a mechanism to drive the upflows. \citet{Murray2010} performed a 3D MHD simulation, and they found that an upward acceleration of the coronal plasma is achieved when an AR expands horizontally in a unipolar background-field environment. They have managed to reproduce upflow speeds up to 45\,km\,s$^{-1}$. Since AR expansion naturally leads to an intensification of the electric current at the interface between closed- and open-field structures, which favors the occurrence of magnetic reconnection, it is likely that reconnection and expansion both contribute to the generation of the AR upflows. This was demonstrated when upflow velocities are significantly enhanced on the side of an AR adjacent to the open field of a nearby coronal hole. The slow rise and expansion of a flux rope contained within the AR lead to stronger compression of the open field and the intensification of upflows hours before the flux rope erupts as a CME \citep{Baker2012}. Using a similar approach, \citet{Harra2012b} found that the upflows at the west and east sides of an emerging AR are dominated by reconnection jets and pressure-driven upflows, respectively. 

\citet{Hara2008} conjectured that high-speed upflows inferred from the blueward asymmetries of coronal line profiles result from impulsive heating at the base of the corona, around the footpoints of AR loops. Based on an observed association of the episodic high-speed upflows with chromospheric activity, \citet{McIntosh2009b} suggested that these upflows represent discrete mass-injection events produced by episodic local heating of the chromosphere. A similar scenario was also proposed based on the similar velocity distributions of these high-speed coronal upflows and fast chromospheric spicules \citep{DePontieu2009}. In limb observations, some chromospheric spicules appear to be heated to TR and coronal temperatures, exhibiting as upward propagating upflows in the AIA EUV passbands \citep{DePontieu2011}. Based on these observations, these authors proposed that the high-speed coronal upflows are actually heating signatures during the upward propagation of spicule-like plasma jets that are possibly driven by magnetic reconnection \citep[e.g.\,][]{DePontieu2009,Samanta2019,ChenY2019} or other processes such as amplified magnetic tension \citep{Martinez-Sykora2017} in the chromosphere. They further pointed out that these episodic high-speed upflows generated at chromospheric heights play a key role in the mass and energy supply to the corona. 

Although this scenario has received a lot of attention in the community, it has also been questioned by some studies. For instance, \citet{Klimchuk2012} examined a scenario of spicule-supplied coronal plasma, which predicted a much larger blue-wing enhancement than observed. They concluded that spicules cannot provide sufficient pre-heated plasma to fill the corona. Even if they did, additional heating would be needed to maintain the high temperature as the plasma expands upward and cools adiabatically. \citet{Klimchuk2014} and \citet{Bradshaw2015} performed one-dimensional hydrodynamic simulations for the idea of coronal heating by spicules, and they found that the synthesized coronal spectral-line profiles are distinctly different from the observed ones. They concluded that impulsive heating events in the chromosphere cannot explain the bulk of coronal heating, although they may be responsible for heating of the chromospheric spicules to TR temperatures. With EIS observations of multiple spectral lines formed at different temperatures, \citet{Tripathi2013} derived the emission-measure distribution for the secondary component,  and they concluded that the emission measure is too small to support the proposal of coronal mass supply by chromospheric spicules. Using the density sensitive line pair Fe\,{\sc{xiv}}\,264.78\,{\AA}/274.20\,{\AA}, \citet{Patsourakos2014} found similar densities for the primary and secondary components at most pixels, which agrees with the prediction from a scenario of coronal mass supply through chromospheric evaporation driven by coronal nanoflares. Plasma composition measurements (see below) of the high-speed component by \cite{Brooks2012} also support this scenario. This conclusion finds some support from observations by \citet{Vanninathan2015}, who found no evidence of a spicule contribution to the AR upflows from simultaneous coronal and chromospheric observations, although their study was restricted to looking at asymmetries in the H$\alpha$ line profile, the formation of which is highly complex. In contrast, a recent study by \citet{Polito2020} has found clear evidence of a correlation between the coronal upflows and spectroscopic signatures in TR and chromospheric lines observed with IRIS. The C\,{\sc{ii}}\,1335.71\,{\AA} line is marginally blue-shifted in the upflows compared to in the AR core, Si\,{\sc{iv}}\,1393.75\,{\AA} is less red-shifted, and Mg\,{\sc{ii}}\,k2\,shows a positive asymmetry, which may be interpreted as signatures of upflows. 

By examining the evolution of underlying photospheric magnetic field, \citet{Su2012} identified clear signatures of flux cancellation around one footpoint of an AR loop. Fast jet-like upflows were found to initiate from the footpoint, and they are accompanied by transient brightenings in EUV images. However, no evident signature was found in images of the low chromosphere. They suggested that these upflows result from magnetic reconnection at the height of the upper chromosphere. \citet{Liu2014} found similar results, and they proposed that the intermittent upflows are produced by reconnection between small-scale emerging bipoles and pre-existing open-field structures at AR boundaries.

Recent observations by the \textit{Parker Solar Probe} \citep[PSP:][]{Fox2016} have revealed quasi-periodic Type-III radio bursts that are well correlated with coronal intensity variations at the footpoints of AR loops, suggesting electron acceleration during impulsive reconnection process around the loop footpoints \citep{Cattell2020}. It is unclear whether these radio bursts are signatures of the reconnection processes that generate the AR upflows, but Harra et al. (2021) have found that a radio noise storm during PSP Encounter 2 (between 31 March and 4 April 2019) likely originates in the expansion of the upflow region at the boundary of the only AR on the visible disk during that period.

\subsection{Connection to the Solar Wind} 
  \label{wind}

  \begin{figure}    
   \centerline{\includegraphics[width=0.99\textwidth,clip=]{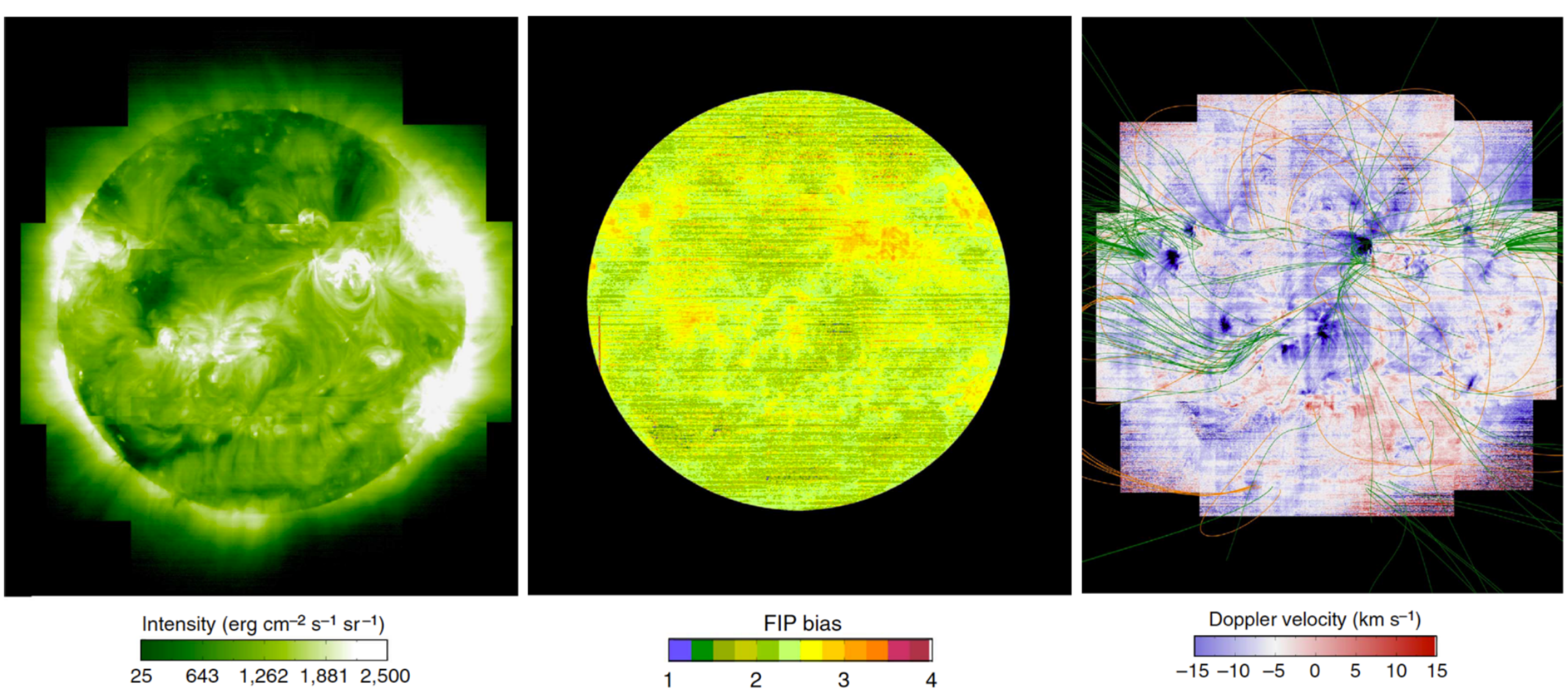}
              }
              \caption{Source regions of the slow solar wind \citep{Brooks2015}. From left to right: full-Sun images of the Fe\,{\sc{xiii}}\,202.04\,{\AA}\,line intensity, Si/S FIP bias and Fe\,{\sc{xiii}}\,202.04\,{\AA}\,Doppler shift. Some open (green) and closed (orange) field lines are overlaid on the Dopplergram. }
   \label{fig_ar_wind}
   \end{figure}
   
The upflows are generally found at the edges of ARs, where EUV spectral lines formed at typical coronal temperatures often show reduced emission. The electron density and temperature there are both lower than those in the cores of ARs \citep{Doschek2008}. Interestingly, coronal holes also show reduced coronal emission, lower electron density, and lower temperature compared to neighboring quiet-Sun regions. These similarities between AR edges and coronal holes suggest that AR boundaries might also be source regions of the solar wind. Indeed, \citet{Kojima1999} found that low-speed solar-wind streams appear to be associated with regions of large magnetic-flux expansion in the vicinity of ARs. This association was confirmed by \citet{Ko2006}, who compared the coronal electron temperature and abundances in the vicinity of an AR with in-situ measurements of the slow solar wind. 

Magnetic-field extrapolations have been performed to understand the magnetic-field structures associated with the persistent upflows. Although a few investigations showed that a portion of the upflows may be guided by large-scale magnetic loops and eventually flow downward to the other footpoints of the loops \citep[e.g.\,][]{Boutry2012}, most studies clearly demonstrated a close association of at least some of these upflows with open-field lines originating from AR boundaries \citep[e.g.\,][]{Marsch2004,Sakao2007,Harra2008,Marsch2008,Baker2009}. \citet{vanDriel-Gesztelyi2012} performed both local magnetic-field extrapolations for individual ARs and global potential-field source-surface modeling. They found that a part of the AR upflows are confined in closed-field structures, and that the other part of the upflows are associated with field lines extending to the source surface from a coronal null point. This association strongly suggests that these AR upflows likely flow outward into interplanetary space and become part of the slow solar wind. However, \citet{Culhane2014} found that their studied AR was completely covered by the closed field lines below the source surface, while the in-situ measurements, combined with the back-mapping technique, strongly suggest the origin of a slow-wind stream from the vicinity of this AR. To solve this apparent inconsistency, \citet{Mandrini2014} proposed a two-step reconnection process. After the first reconnection between the expanding AR loops and neighboring large-scale loops, new large-scale loops connecting the AR upflow and a distant location are formed. These loops then reconnect with open-field lines from a coronal hole around a coronal null point, releasing plasma of the AR upflow into interplanetary space. \citet{Harra2017} have also shown that the upflowing plasma could be released through interchange reconnection if the upflow-hosting ARs are close to an open-field region.  

Chemical composition could be used to establish a link between some structures in the interplanetary solar wind and their source regions in the solar corona \citep[e.g.][]{Feldman2003,Song2020}. Chemical elements with a first ionization potential (FIP) below and above $\approx$10\,eV are often called low-FIP and high-FIP elements, respectively. It is well-known that the relative abundance of a low-FIP element to a high-FIP element is generally enhanced in the slow solar wind relative to its photospheric value by a factor of three to four. This factor is called the FIP bias. Spectroscopic observations at EUV wavelengths can be used to measure the FIP bias in the solar corona. Based on \textit{Hinode}/EIS observations, \citet{Brooks2011} found that the ratio of low-FIP Si and high-FIP S is enhanced by a factor of three to four at the locations of AR upflows. A similar ratio was found in the in-situ measurements of solar wind a few days later, providing evidence for a connection between the observed solar-wind stream and the AR upflows.  In a following study, \citet{Brooks2015} applied a similar technique to the EIS data obtained through a full-Sun mosaic observation campaign and obtained a full-Sun composition (FIP bias) map. After comparing this composition map with the simultaneously obtained full-Sun Dopplergram and the global magnetic-field topology, they identified multiple locations of AR upflows as source regions of the slow solar wind (Figure\,\ref{fig_ar_wind}). 

Using the technique of double-Gaussian fit, \citet{Brooks2012} decomposed the two components of the coronal line profiles observed at the boundaries of an AR. They found that the FIP bias for the high-speed upflow component is also in the range of three to five, suggesting that the high-speed component may also contribute to the slow solar wind. The complexity of the outflow components contributing to the solar wind was also highlighted by \cite{Brooks2020}, who used \textit{Hi-C} \citep{Rachmeler2019} high spatial resolution images at 172\,\AA\ to try to isolate the upflows cleanly from the cooler fan loops. Their results suggest that the variability in solar-wind composition measurements might be explained by activity in the source region.

By applying the technique of Doppler dimming to the spectra obtained with SOHO/UVCS in the height range of 1.5 to 2.3 solar radii from the solar center, \citet{Zangrilli2012} and \citet{Zangrilli2016} identified coronal outflows at the edges of ARs, and they found an increase of the outflow speed from $\approx$50\,km\,s$^{-1}$ at 1.5 solar radii to $\approx$150 km\,s$^{-1}$ at 2.5 solar radii. Since the speeds at 1.5 solar radii are generally smaller than those of the secondary components in EIS observations, it is possible that the fast upflows corresponding to the secondary components mostly supply mass to the corona and do not directly escape to the interplanetary space. Instead, the primary components in EIS observations, with a blue shift of about 10\,km\,s$^{-1}$ \citep[e.g.\,][]{Tian2011b}, may be the major source of the outflows detected by UVCS. It is also possible that what UVCS measured is an average speed, meaning that the weak and transient high-speed upflows may have been smeared out over the course of temporal or spatial sampling.   

\begin{figure}    
   \centerline{\includegraphics[width=0.99\textwidth,clip=]{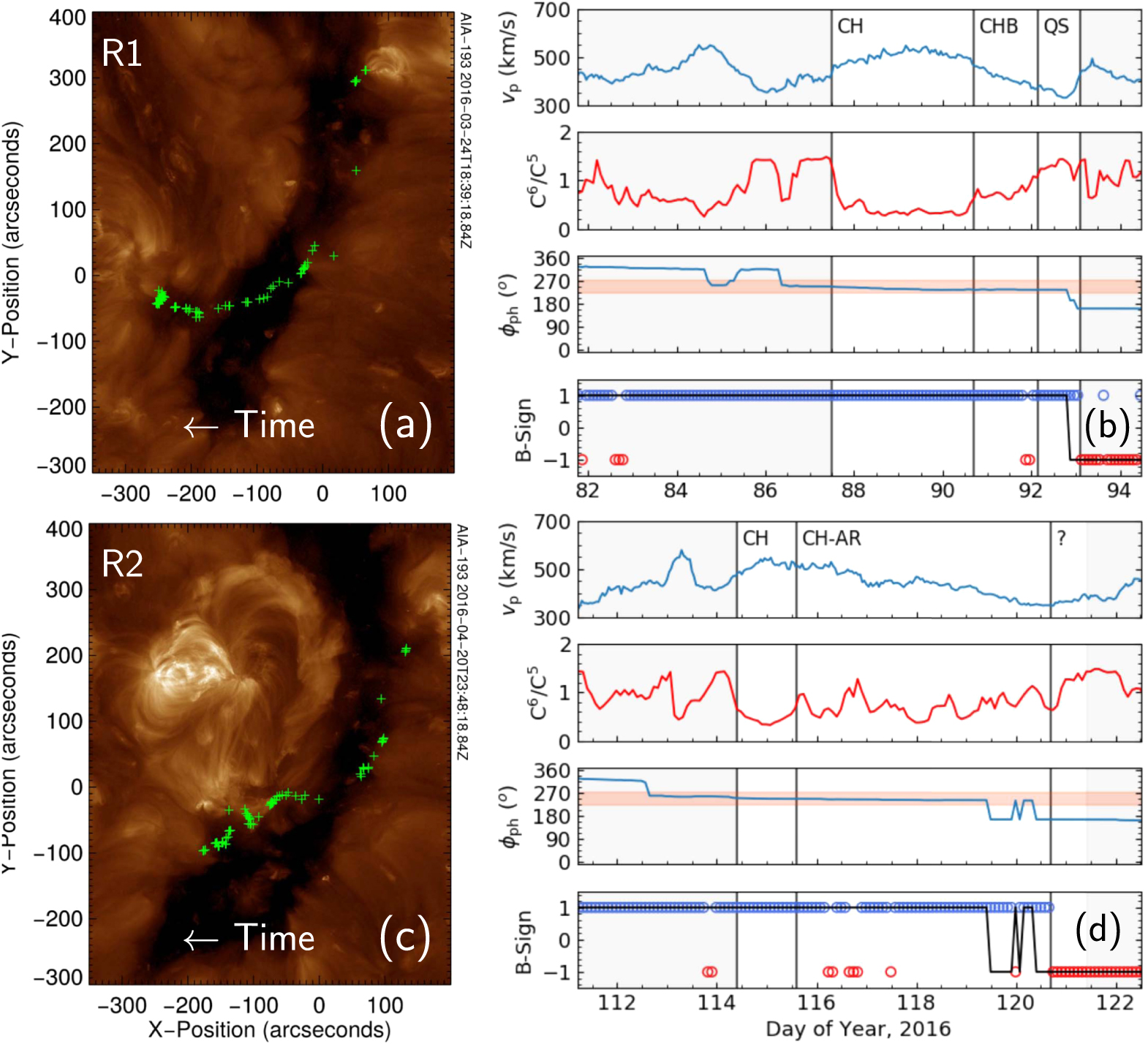}
              }
              \caption{ Left: AIA 193\,{\AA} images taken during two solar rotations \citep{Macneil2019}. The solar wind source points are overplotted on the images. Right: time series of several solar-wind parameters, with associated source regions labeled and separated by vertical lines. In panels b or d the first two rows correspond to the solar-wind velocity and C$^{6+}$/C$^{5+}$, respectively. The third row shows the sourcepoint longitude for each mapped data point. The orange bar indicates longitudes corresponding to the vicinity of the CH. The fourth row shows the magnetic-field polarity of the radial component of the interplanetary magnetic field (circles) and the corresponding PFSS magnetic-field polarity for each mapped data point (black line). Reproduced by permission of the AAS. }
   \label{fig_ar_linkage}
   \end{figure}
   
Longer-term measurements are also helpful for understanding the linkage between the upflows and the solar wind. By comparing the synoptic photospheric magnetograms and the time sequences of solar-wind parameters, \citet{He2010} concluded that the upflowing plasma at an AR boundary may evolve into an intermediate-speed solar-wind stream observed near the Earth. This solar-wind stream has a speed of about 400\,km\,s$^{-1}$, and an intermediate temperature and density if comparing to the typical fast and slow solar wind. In another study by \citet{Janardhan2008}, the authors managed to establish a link between an extremely slow ($\leqslant$300\,km\,s$^{-1}$) solar-wind stream measured in interplanetary space and an evolving low-emission coronal region located near an AR. They claimed that interchange reconnection between the AR loops and surrounding open fields leads to a reduction in the area of the open-field region, which would result in a smaller solar-wind velocity. With the help of the potential-field source-surface (PFSS) model, \citet{Liewer2004} ballistically mapped several solar-wind streams back to the source surface and then to the photosphere. They found that these solar-wind streams generally can be traced back to dark areas in EUV and soft X-ray images at the edges of ARs. \cite{Fazakerley2016} tracked a Carrington rotation that had coronal holes and active regions. The solar wind was linked back to the features on the Sun, using ballistic back mapping and PFSS modelling. They found short periods of enhanced-velocity solar wind at the boundary of slow and fast wind streams, which are related to ARs that are located beside coronal holes. The neighbours of ARs are important in terms of how the solar wind is created. Figure\,\ref{fig_ar_linkage} shows an example of connecting remote sensing data that are showing regions of upflows in a coronal hole and an AR, with the solar-wind measurements \citep{Macneil2019}. These authors examined two solar rotations, one with the quiet Sun and coronal hole only, and during the next rotation an AR has emerged, making it ideal to compare the two different situations. Their observations indicate that the features that emerge in the AR-associated wind are consistent with an increased occurrence of interchange reconnection during solar-wind production, compared with the initial quiet-Sun case. Back-mapping the solar wind to the source surface to trace solar wind source regions has also been trialed on \textit{Hinode} and ACE observations as a science preparation for \textit{Solar Orbiter} \citep{Stansby2020}, and the authors discussed some of the difficulties involved in interpreting the composition measurements. A direct link between solar-wind streams and their coronal source regions may also be achieved in a different way by using the Wang\,--\,Sheeley\,--\,Arge (WSA) model \citep{Wang1990,Arge2000}. Using the WSA model, \citet{Slemzin2013} have also identified locations of AR upflows as the source regions of slow solar-wind streams. 

Tracing the solar wind back to the Sun from 1999 to 2008, \cite{Fu2015} classified the solar wind by its source-region types with the EUV images and photospheric magnetograms taken by SOHO. They found that about half of the solar-wind streams come from AR regions during the solar maximum. The charge states, FIP bias and helium abundance are significantly different for the solar wind coming from ARs, the quiet Sun, and coronal-hole regions \citep{Fu2017,Fu2018}. Their results further indicate that the reconnection between the open magnetic-field lines and closed loops may play a major role in the generation of the AR-associated wind.

\section{Upflows from CME-Induced Dimmings} 
      \label{dimming} 

When coronal mass ejections (CMEs) leave the Sun, often we see reduced emission of the corona, most noticeably at EUV and soft X-ray wavelengths \citep[e.g.\,][]{Hudson1996,Sterling1997,Dissauer2018,Dissauer2019}. This phenomenon is called coronal dimming, and a dimming region is sometimes called a transient coronal hole. Statistical studies show that more than half of the frontside CMEs are associated with dimmings \citep{Reinard2008,Bewsher2008}. There are at least two types of coronal dimmings: core (or twin) dimmings and secondary dimmings. Core dimmings usually refer to small regions of greatly reduced coronal emission near the eruption sites of CMEs. They are generally believed to mark the footpoints of ejected flux ropes, and they could last for more than ten hours \citep[e.g.\,][]{Hudson1996,Vanninathan2018,ChenH2019,Xing2020}. Secondary dimmings, which may result from stretching of the magnetic-field lines or reconnection of the erupting magnetic field with the surrounding structures, often show a less prominent intensity decrease and a quicker recovery \citep[e.g.\,][]{Thompson2000,Mandrini2007,Zheng2016,Vanninathan2018}. Note that some dimmings are likely not associated with CMEs, e.g. long-duration remote dimmings associated with confined circular-ribbon flares \citep{ZhangQ2020} and dimmings at the peripheries of emerging flux regions \citep{ZhangJ2012}. Here we focus on CME-induced dimmings. Spectroscopic observations often reveal an obvious density decrease in these dimming regions \citep{Harrison2000, Harrison2003,Tian2012a}. In addition, the peak temperature of the differential-emission-measure curve appears not to change when a dimming occurs \citep{Tian2012b}. These results demonstrate that coronal dimming is mainly due to mass loss rather than temperature change. Signatures of coronal dimming are also present in the Sun-as-a-star spectra taken by the \textit{EUV Variability Experiment} \citep[EVE:][]{Woods2012} onboard SDO \citep{Mason2014}.

  \begin{figure}    
   \centerline{\includegraphics[width=0.99\textwidth,clip=]{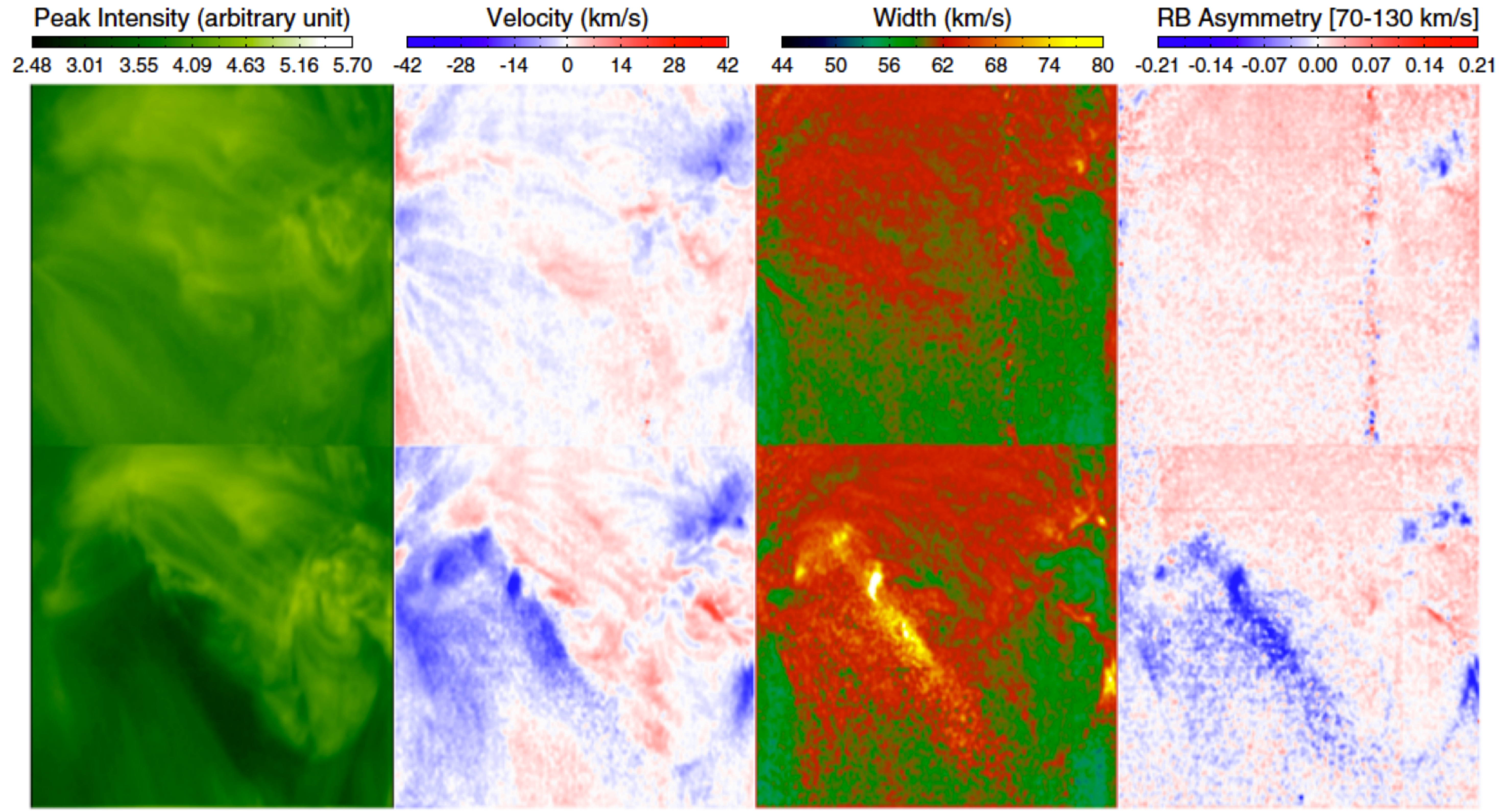}
              }
              \caption{Maps of the peak intensity, Doppler shift, and line width derived from a SGF, and RB asymmetry in the velocity interval of 70\,--\,130 km\,s$^{-1}$ for the Fe\,{\sc{xiii}}\,202.04\,{\AA}\,line \citep{Tian2012a}. The first and second rows show observational results before and after a CME eruption, respectively. The FOV of each image is about 250$^{\prime\prime}$$\times$250$^{\prime\prime}$. The same dataset has also been analyzed by \citet{Harra2007} and \citet{Jin2009}. Reproduced by permission of the AAS. }
   \label{fig_dimming_map}
   \end{figure}
   
There are only a few spectroscopic investigations of coronal dimmings. Using CDS observations, \citet{Harra2001} reported an obvious blue shift in coronal and TR lines in coronal dimming regions. At least a few coronal dimming events have been caught by the EIS spectrograph. These observations often reveal a prominent blue shift of coronal emission lines in dimming regions, normally in the range of $\approx$10\,--\,50\,km\,s$^{-1}$, with an average of about 20\,km\,s$^{-1}$ when a SGF is performed \citep{Harra2007,Harra2011}. These blue shifts appear to be very similar to those at AR boundaries, and they likely represent systematic plasma upflows along the field lines opened by CMEs. The strongest blue shifts are found at the footpoints of disrupted coronal loops \citep{Harra2007,Attrill2010}. The blue shift appears to be positively correlated with the depression of coronal emission and magnetic-field strength in the photosphere \citep{Jin2009}. In addition, the spectral-line profiles are found to be broadened compared to the line profiles in the pre-eruption phase and outside dimming regions. The large line width has been interpreted as a result of amplified Alfv\'en wave amplitude \citep{McIntosh2009c} or inhomogeneity of flow velocities along the LOS \citep{Dolla2011}.

Similarly, the coronal line profiles have been found to be asymmetric in dimming regions. Using the technique of RB-asymmetry analysis introduced by \citet{DePontieu2009}, \citet{McIntosh2010} found a couple of small patches in the dimming region behind an erupted halo CME where the Fe\,{\sc{xiii}}\,202.04\,{\AA}\,line profile is slightly enhanced at the blue wing. Note that they used the SGF to determine the line centroid when performing the RB-asymmetry analysis, which tends to underestimate the degree of the actual profile asymmetry \citep{Tian2011b}. As demonstrated by \citet{Tian2011b}, the secondary component can be more accurately resolved if the wavelength position of the peak spectral intensity is used to determine the line centroid. \citet{Tian2012a} applied this improved RB-asymmetry technique to the same dataset, and found much more prominent and pervasive blueward asymmetries in dimming regions. As shown in Figure\,\ref{fig_dimming_map}, when a dimming occurs, we see not only the intensity depression, obvious blue shift, and enhanced line width derived from a SGF, but also clear blueward asymmetry. The strongest blueward asymmetries appear in areas of strongest intensity reduction. Detailed analyses of the line profiles suggest that a single Gaussian function normally cannot fit the observed line profiles in dimming regions well. Instead, a double-Gaussian fit appears to approximate the observed line profiles very well. So, similar to AR boundaries, the coronal emission in dimming regions also consists of at least two emission components: a nearly stationary background component and a weak high-speed upflow component. The secondary component has an upward speed of about 100\,km\,s$^{-1}$. The intensity ratio of the two components is often about 5\,--\,15\,\%, but at some pixels it could be as high as 40\,\%. \citet{Tian2012a} also found that the SGF blue shift, line width, and blueward asymmetry are all strongly correlated, indicating that the $\approx$20\,km\,s$^{-1}$ blue shift and enhanced line width derived from a SGF are caused at least partly by the superposition of the two components. So it appears clear that a fraction of the coronal plasma in dimming regions flows upward at a speed of the order of $\approx$100\,km\,s$^{-1}$. Recent EUV-imaging observations with AIA have clearly revealed upflows with speeds of $\approx$70\,--\,140\,km\,s$^{-1}$ from a CME footpoint in a dimming region \citep{Lorincik2020}, confirming the finding of \citet{Tian2012a}.

  \begin{figure}    
   \centerline{\includegraphics[width=0.99\textwidth,clip=]{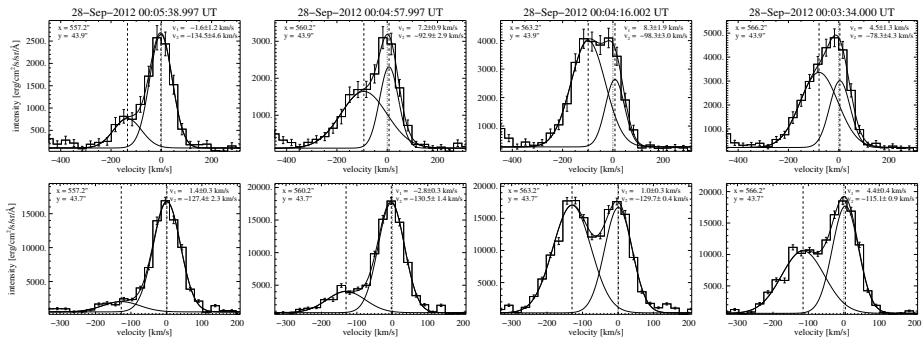}
              }
              \caption{Line profiles of Fe\,{\sc{xiii}}\,202.04\,\AA{}(upper panels) and Fe\,{\sc{xv}}\,284.16\,\AA{}(lower panels) at four pixels within a dimming region \citep{Veronig2019}. The three smooth curves in each panel represent the two Gaussian components and their sum. Reproduced by permission of the AAS. }
   \label{fig_dimming_profile}
   \end{figure} 
   
Similar blueward asymmetry and a high-speed upflow component have been commonly observed in other dimming events. \citet{Chen2010} reported an expanding dimming event with a leading edge/ridge of enhanced line width. \citet{Tian2012a} examined the asymmetry of the coronal spectra and found clear blueward asymmetries along the whole ridge and in the obvious dimming regions behind it. In a recent study, \citet{Veronig2019} found that at the growing dimming border the upflow component is so strong that the spectra even show distinct double components (Figure\,\ref{fig_dimming_profile}). At some pixels the upflow component even dominates over the nearly stationary component. Such line profiles can be decomposed with a high degree of accuracy through a double-Gaussian fit. 

What is the significance of these upflows? Dimming regions are often called transient coronal holes, as the coronal magnetic-field lines are transiently opened by the CME ejecta. So a large fraction of these upflows likely flow outward along the opened field lines and eventually become solar-wind streams following the CME ejecta. In this sense, a dimming region should be regarded as a source region of the solar wind, similar to AR boundaries and coronal holes \citep{McIntosh2009c,Tian2012a}. The enhanced nonthermal broadening may be contributed by both the superposition of different emission components and the growing Alfv\'en wave amplitude in the open-field and low-density coronal environment. The momentum flux resulting from the high-speed upflows may act as a secondary momentum source, and thus it may impact the kinematics of the associated CMEs \citep{McIntosh2010}. It is also possible that these upflows continuously feed the associated CMEs \citep{Harra2007}. Using the density sensitive line pair Fe\,{\sc{xiii}}\,196.55\,\AA{}/202.04\,\AA{}, \citet{Veronig2019} derived an electron density of 2.25$\times$10$^9$\,cm$^{-3}$ for the upflow component. They then estimated a mass-loss rate due to the upflows, which is about 8$\times$10$^{11}$\,g\,s$^{-1}$ and comparable to the mass-increase rate of the associated CME inferred from coronagraph observations. Thus, the upflows may become part of the CME. In addition, since a dimming region will eventually recover to the pre-eruption state, part of these upflows might provide mass to refill the corona after the eruption of CMEs \citep{Tian2012a}. 

Besides these prevalent high-speed upflows that appear to show no obvious temperature dependence, clear temperature-dependent upflows associated with dimmings have also been identified. For instance, \citet{Imada2007} identified an upflow with a speed increasing from $\approx$15\,km\,s$^{-1}$ at 10$^{4.9}$\,K to $\approx$160\,km\,s$^{-1}$ at 10$^{6.3}$\,K at the edge of a dimming region. With the RB-asymmetry technique, \citet{Tian2012a} detected several more temperature-dependent upflows, but all in very small areas immediately outside the (deepest) dimming regions. These temperature-dependent upflows are likely evaporation flows induced by interaction between the stretching field lines in dimming regions and adjacent magnetic-field structures. 

Dimmings have been linked to magnetic-cloud materials. \citet{Harra2011} used upflow measurements to determine accurately where the dimming occurred through a velocity-difference method. They used this to determine the maximum amount of magnetic field that could end up in a magnetic cloud. The magnetic-cloud measurements and modelling determined that the magnetic flux in the dimming region is consistent with the magnetic-cloud data.

\section{Summary and Future Perspectives} 
      \label{summary} 
 
In the past two to three decades, EUV/FUV spectroscopic observations have largely changed our view of the upper solar atmosphere. With these dedicated observations, highly structured upflows and downflows with varying magnitudes at different temperatures have been identified in various regions on the Sun. On large scales these flows appear to be quasi-steady and long-lasting, possibly indicating a continuous global plasma circulation \citep{Marsch2008}. Systematic upflows in the upper TR and corona have been commonly believed to be signatures of the nascent solar wind. High-cadence observations and detailed analyses of the spectral-line profiles suggest that there are intermittent high-speed upflow components with a wide range of temperatures and a relatively slow cooling downflow component mostly at TR temperatures. These observations indicate that the upper atmosphere is not a static and magnetically stratified layer, but rather a dynamic interface revealing a continuous mass cycling between the chromosphere and corona/solar wind.

We have presented an extended overview of plasma flows observed by the EUV/FUV spectrographs of \textit{Hinode}/EIS, IRIS, SOHO/SUMER, and SOHO/CDS, with an emphasis on the upflows seen in the network structures of the quiet Sun and coronal holes, boundaries of ARs, and CME-induced dimming regions. Despite significant advances in the research of these plasma flows, several important unresolved issues remain and require further investigations.

For the quiet Sun and coronal holes, there is still no consensus on the physical mechanisms behind the observed temperature dependence of the Doppler shift. It is commonly believed that this temperature dependency is a consequence of coronal heating \citep{Peter2006,Hansteen2010}. So more realistic numerical simulations of coronal heating should be performed to better understand this temperature dependency. In addition, the exact temperature at which the net Doppler shift changes sign is still unclear, mainly because there are no strong emission lines formed between the redshift-dominated temperatures of  $\log\,T\leqslant 5.4$ and the blueshift-dominated temperatures of $\log\,T\geqslant 5.8$ in the SUMER spectral range. Considering this, the strong Ne\,{\sc{vii}}\,465\,\AA{}\,line formed around $\log\,T=5.7$ is highly recommended for future EUV instruments aiming at high-resolution observations of the quiet Sun and coronal holes \citep{Tian2017}.

For the upflows at AR boundaries, at least three important questions or issues need to be addressed in the future. First, it is clear now that both flows and waves are present at AR boundaries. However, we still do not know how exactly the upflows and waves are generated and related. So far only a few numerical models exist and they are not fully compatible with observational characteristics. Although a few case studies have shown local heating and injection of individual fast upflows at loop footpoints \citep[e.g.\,][]{Li2019}, detailed observations of the generation, propagation, and heating of individual upflow events are generally missing. Second, the contribution of the fast upflows to coronal heating is still highly debated. Contradictory conclusions have been reached based on different approaches. Third, although a few studies have managed to trace several slow solar-wind streams back to the coronal sources, a one-to-one connection between coronal upflows and interplanetary solar-wind streams still cannot be established routinely. In the future, more advanced MHD simulations should be performed to understand the generation, propagation, and energization of the upflows and waves. In addition, the \textit{Spectral Imaging of the Coronal Environment} \citep[SPICE:][]{Spice2020} instrument onboard the newly launched \textit{Solar Orbiter} mission \citep{Mueller2020} has the capability of sampling several strong emission lines formed at both TR and coronal temperatures at a high spatial and temporal resolution. Combined observations between SPICE and the 4-m \textit{Daniel K. Inouye Solar Telescope} (DKIST) will likely reveal new insights into the generation and propagation of the upflows/waves as well as their role in the energization of the coronal plasma. Having a mission such as \textit{Solar Orbiter} in an orbit away from the Earth's orbit will also allow 3D spectroscopy by combining \textit{Solar Orbiter}/SPICE with the Earth-orbiting \textit{Hinode}/EIS and IRIS missions. A main goal of the \textit{Solar Orbiter} mission is to trace the interplanetary solar-wind streams back to their sources in the corona more accurately and routinely by combining imaging and spectroscopic observations with in-situ measurements. FIP bias measurements both spectroscopically and in situ will form a significant part of this work. New insight could also be obtained through observations with a future spectrometer, the \textit{Solar-C EUV High-Throughput Spectroscopic Telescope} \citep[EUVST,][]{Shimizu2019}, which has been approved by JAXA. \textit{Solar-C}/EUVST will provide near continuous spectral measurements throughout the solar atmosphere at similar spatial resolutions. The spatial resolution will be around seven times better than that of \textit{Hinode}/EIS, and the temperature coverage allows us to observe seamlessly from the chromosphere ($\approx$ 10,000\,Kelvin), to the corona ($\approx$ 1\,million Kelvin), and flare plasmas ($\approx$ 10\,million kelvin). Normally magnetic-field extrapolations are required to guide our understanding of the corona--solar-wind connection. With recently developed promising techniques of coronal magnetic-field measurements \citep{Yang2020a,Yang2020b,Li2015,Li2016,Si2020,Landi2020}, magnetic-field extrapolations will likely improve and thus help establish a more accurate connection between the coronal upflows and interplanetary solar-wind streams. In addition, the coronal magnetic field measurements from DKIST will allow us to fully understand the coronal magnetic field rather than rely completely on models. These new facilities will revolutionize our understanding of the energy transport in the solar atmosphere \citep{Velli2020}. 

Formation mechanisms of the upflows in CME-induced dimming regions, and their potential role in solar-wind formation and impact on CME evolution, are still poorly understood. This is mainly due to the infrequent spectroscopic observations of dimmings because of the small fields of view and lower cadence of the slit scans. Possibly, future instrumentation should aim at full-disk spectroscopic imaging with a wide temperature coverage to catch a large number of dimming events. In the \textit{Solar Orbiter} era, we also expect that the linkage between dimming regions and solar-wind streams will be made more easily than before.

\begin{acks}
H. Tian is supported by NSFC grants 11825301 and 11790304. The work of D. H. Brooks was performed under contract to the Naval Research Laboratory and was funded by the NASA Hinode program. D. Baker is funded under STFC consolidated grant number ST/S000240/1. L. Xia is supported by NSFC grants 41974201 and 41627806. H. Tian acknowledges support from the UCL-PKU strategic partner funds during his visit to MSSL. This article is based upon the AAS/SPD Karen Harvey Prize Lecture of 2020, the presentation file of which is available at \url{http://spd.aas.org/prizes/harvey/previous}. 

\end{acks}

{\footnotesize\paragraph*{Disclosure of Potential Conflicts of Interest}
The authors declare that they have no conflicts of interest. 
}


\bibliographystyle{spr-mp-sola}
\bibliography{upflow_bib.bib}

\end{article} 

\end{document}